\documentclass[useAMS,usenatbib]{mn2e}
\usepackage{graphicx}
\usepackage{multirow}
\usepackage{afterpage}
\usepackage{lscape}
\usepackage{amsmath}
\usepackage{longtable}
\setlongtables

\def\gtrsim{\mathrel{\hbox{\rlap{\hbox{\lower4pt\hbox{$\sim$}}}\hbox{$>$}}}}
\def\lesssim{\mathrel{\hbox{\rlap{\hbox{\lower4pt\hbox{$\sim$}}}\hbox{$<$}}}}
\def\apjs{{ApJS}}
\def\mnras{{MNRAS}}
\def\apj{{ApJ}}
\def\aap{{A\&A}}
\def\aj{{AJ}}
\def\araa{{ARA\&A}}
\def\apjl{{ApJ}}
\def\nat{{Nature}}
\def\aaps{{A\&AS}}
\def\pasp{{PASP}}
\def\baas{{BAAS}}%

\title[The distribution of AGN in clusters of galaxies: Paper I]{The distribution of AGN in a large sample of galaxy clusters }
\author[R. Gilmour, P. Best, and O. Almaini]{R. Gilmour$^{1,2}$\thanks{e-mail rgilmour@eso.org}, P. Best$^{2}$ and O. Almaini$^3$\\
$^{1}$European Southern Observatory, Alonso de Cordova 3107, Vitacura, Casilla 19001, Santiago 19, Chile\\
$^{2}$Scottish Universities Physics Alliance, Institute for Astronomy, Royal Observatory, Blackford Hill, Edinburgh, EH9 3HJ, UK\\
$^{3}$School of Physics and Astronomy, University of Nottingham, University Park, Nottingham, NG7 2RD, UK}
\begin{document}

\date{Accepted . Received ; in original form }

\pagerange{\pageref{firstpage}--\pageref{lastpage}} \pubyear{2002}

\maketitle

\label{firstpage}

\begin{abstract}

We present an analysis of the X-ray point source populations in 182
{\it Chandra} images of galaxy clusters at $z>0.1$ with exposure time
$>$10 ksec, as well as 44 non-cluster fields.  Analysis of the number
and flux of these sources, using a detailed pipeline to predict the
distribution of non-cluster sources in each field, reveals an excess
of X-ray point sources associated with the galaxy clusters. A sample
of 148 galaxy clusters at $0.1<z<0.9$, with no other nearby clusters,
show an excess of 230 cluster sources in total, an average of
$\sim$1.5 sources per cluster. The lack of optical data for these
clusters limits the physical interpretation of this result, as we
cannot calculate the fraction of cluster galaxies hosting
X-ray sources. However, the fluxes of the excess sources indicate that
over half of them are very likely to be AGN, and the radial
distribution shows that they are quite evenly distributed over the
central 1 Mpc of the cluster, with almost no sources found beyond this
radius.  We also use this pipeline to successfully reproduce the
results of previous studies, particularly the higher density of
sources in the central 0.5 Mpc of a few cluster fields, but show that
these conclusions are not generally valid for this larger sample of
clusters. We conclude that some of these differences may be due to the
sample properties, such as the size and redshift of the clusters
studied, or a lack of publications for cluster fields with no excess
sources. This paper also presents the basic X-ray properties of the
galaxy clusters, and in subsequent papers in this series the
dependence of the AGN population on these cluster properties will be
evaluated.

In addition the properties of over 9500 X-ray point sources in the
fields of galaxy clusters are tabulated in a separate catalogue
available online or at www.sc.eso.org/$\sim$rgilmour.

\end{abstract}


\section{Introduction}

Studies of AGN host galaxies are key to understanding the physical
mechanisms which trigger AGN activity, and govern the fueling rate of
the central black hole. An important part of these studies is the
external environment of the galaxies, which has long been known to
have a significant link with the galaxy properties (e.g
\citealt{Hubble}). Correlations such as the morphology--density
\citep{DresslerMorph} and star-formation -- density
(e.g. \citealt{SLOANcluster}) relations are evidence of the significant
transformations that are associated with galaxy clusters. If AGN
activity is influenced by, for example, galaxy mergers or
gravitational disruption of the host galaxy, then the number of AGN
would also differ between galaxy clusters and the field, and within
the cluster itself.

The first evidence of a link between AGN activity and environment came
in 1978, when \cite{Gisler78} found a lack of emission-line galaxies
in galaxy clusters relative to the field, which was confirmed by
Dressler, Thompson \& Shectman (1984) \nocite{DresslerSurvey}. More
recently, large optical surveys have been used to identify the
properties of a significant number of host galaxies, and hence compare
AGN activity in the same type of host galaxy but different
environment. \cite{Miller} find that optical AGN activity is
independent of environment, but \cite{Wake04} show that the level of
clustering depends on AGN luminosity.  \cite{Kauffmann} find that AGN
with strong [OIII] emission avoid areas of high galaxy
density. \cite{Best} investigate the radio properties of these AGN,
and find that the fraction of galaxies with radio-loud AGN increases
dramatically with local galaxy density, but that all of these AGN have
low [OIII] emission and so may not be seen in optical surveys.

Arguably the least biased method of detecting AGN currently is to use
X-ray images, which have the added advantage that the vast majority of
point sources are AGN.  Not long after optical surveys identified a
lack of emission-line galaxies in clusters, X-ray surveys began to
find a surprisingly high number of point sources in fields with galaxy
clusters (\citealt{Bechtold}; \citealt{Henry};
\citealt{Lazzati}). With the advent of the {\it Chandra} X-ray telescope,
with sub-arcsecond point sources, such studies were repeated for other
clusters, with a range of results.  Significant overdensities of point
sources have been found in a number of fields with galaxy clusters at
moderate redshifts (\citealt{Cappi}, z=0.5; \citealt{Martini}, z=0.15;
\citealt{Molnar}, z=0.32; \citealt{Olivia}, z=0.83;
\citealt{Martini2}, $0.05 < z < 0.31$; \citealt{DElia04}, z=0.5) and
groups (\citealt{Jeltema}, $0.2 <z < 0.6$), but \citeauthor{Molnar}
also found a z=0.5 cluster without a significant overdensity. More
recently, studies of significant samples of galaxy clusters by
\cite{Cappelluti} (10 clusters, $0.24<z<1.2$), \cite{Ruderman} (51
clusters, $0.3<z<0.7$), and \cite{Branchesi} (18 clusters,
$0.25<z<1.01$) have all found significant overdensities of point
sources over the full sample, but not necessarily in all individual
fields.  However, the ChaMP project \citep{Champ2} find no difference
in the number density of sources in fields with $z > 0.3$ clusters
compared to those without clusters.

The calculated number of AGN per cluster varies significantly in these
samples, even taking into account the different depths of the
observations and the statistical variance in the number of background
sources in each image. This is to be expected as the number of AGN per
cluster is, of course, related to the cluster properties, such as the
number of possible AGN host galaxies. However, due to the lack of
optical data it is hard to draw any conclusions from these samples as
to how, if at all, the cluster environment affects the AGN
population. Optical imaging and spectroscopy can identify the X-ray
detected AGN in the cluster, rather than relying on statistical
background subtraction, and also reveal the distribution and number of
normal cluster galaxies. This would allow the calculation of the fraction
of cluster galaxies which host AGN, as a function of cluster radius or
cluster size for example, shedding light on the physical mechanisms
affecting AGN. On the other hand obtaining optical data for a large
sample of galaxy clusters is time consuming, and any strong trend
should be visible in the distribution of X-ray point sources in a
large sample of cluster fields. A rough estimate of the cluster galaxy
population can also be made from the extended X-ray gas. An analysis
based purely on X-ray data can therefore be very useful, but is
clearly inferior to a full spectroscopic analysis of a large sample of
clusters, with X-ray and optical data.

Such a study was started by \cite{Martini}, using optical data to
identify X-ray detected AGN in galaxy clusters.  Martini, Mulchaey \&
Kelson (2007)\nocite{Martini_pos} have confirmed optically that the
{\it fraction} of galaxies hosting X-ray detected AGN does differ
significantly between the eight clusters in their sample, implying
that the cluster properties affect the number of AGN. Their
spectroscopic data confirms between 2 and 10 AGN per cluster, with a
range in AGN fractions that cannot be explained by Poissonian
variations. The mean fraction for the whole sample is 5 per cent of
galaxies with $M_R<-20$ hosting AGN with $L_X > 10^{41}$ erg
sec$^{-1}$. The wide variation in the number or fraction of X-ray AGN
is also found when other studies with spectroscopic data are
compared. For example \cite{Finoguenov04} find only one confirmed
X-ray AGN, with luminosity $\sim 10^{41}$ erg sec$^{-1}$, in a 1.8
square degree survey of the centre of the Coma cluster (z=0.02), but
\cite{Davis03} find between 3 and 5 AGN in a cluster at z=0.08, giving
a fraction of 4 per cent in agreement with the mean value of
\citeauthor{Martini_pos}

\citeauthor{Martini_pos} also find tentative evidence that the
clusters with higher AGN fraction have lower redshift, lower velocity
dispersion, higher substructure and lower Butcher--Oemler
\citep{ButcherOemler} fraction, but due to the small size of the
sample (8 clusters) it is not clear from this sample which, if any, of
these factors is affecting the AGN activity. A higher AGN fraction at
high redshift would be expected from the field evolution of AGN, but
no strong evolution is found by \cite{Branchesi} or
\cite{Ruderman}. In contrast, \cite{Cappelluti} find some evidence for
an increase in AGN with redshift, and \cite{Eastman07} conclude that
the increase of bright AGN in clusters is up to 20 times greater than
in the field between $z\sim 0.2$ and $z \sim 0.6$.

If AGN are triggered by galaxy mergers then clusters with lower
velocity dispersions would be expected to have higher AGN fractions,
as they have a higher merger rate. \cite{Popesso06} show that this is
indeed the case for optically detected AGN, with clusters with high
velocity dispersions having lower AGN fractions. However
\cite{Martini_pos} find that the velocity distribution of AGN in eight
clusters is not significantly different from the velocity distribution
of the non-active cluster galaxies.

Galaxy mergers are also more common in the outskirts of clusters, so
the projected radial distribution of AGN should be different from the
host galaxies if mergers cause AGN activity. \citeauthor{Martini_pos}
find no evidence for AGN to lie in galaxies in the outskirts of the
cluster compared to bright cluster galaxies, in fact on the contrary,
they find evidence that galaxies with luminous AGN ($>10^{42}$ erg sec$^{-1}$)
are more centrally clustered than the general
population. \citeauthor{Branchesi} and \citeauthor{Ruderman} also
present evidence that most AGN are found within the central 0.5 Mpc of
the clusters, 
which \citeauthor{Ruderman} attribute to tidal encounters with the
central galaxy. In contrast, \cite{Olivia} find the excess AGN in a
cluster at z=0.83 lie between 1 and 2 Mpc from the cluster
centre. \cite{Gilmour1} also find that AGN avoid the densest areas of
a supercluster at z=0.17.

The wide range of results from the current studies may partly be due
to the large number of variables which can affect the AGN fraction in
clusters, and partly due to statistical fluctuations in the small
number of AGN found per cluster. In addition, studies which count
X-ray point sources, without optical confirmation of cluster
membership, are limited by field-to-field variation in the clustering
of background sources (e.g. \citealt{Gilli}). Careful data reduction
is also required, as detecting AGN against the extended intra-cluster
medium and accounting for the variations in the point spread function
(PSF) is important in determining the expected number and distribution
of background AGN in the field. Different treatments of these
variables may account for the differences between the results of the
ChaMP project and the surveys which deliberately target AGN in cluster
emission. In order to understand how many AGN are in clusters with
different properties, and to get significant statistics, a large
sample of galaxy clusters and non-cluster fields are required.  Even
in a such a sample the conclusions are limited by the lack of optical
data which means that individual cluster AGN cannot be identified.  We
may, however, expect to see any strong trends in the average number of
AGN in clusters of a given type, or at a given epoch.  As shown later
in this paper, at least five fields are required in each subsample to
remove the effects of cosmic variance in the background
distribution. Larger samples are likely required in order to get
statistically significant results.

\section{Outline and Method}

The large number of observations of galaxy clusters in the {\it Chandra}
archive provide an excellent basis for investigating the prevalence of
AGN in galaxy clusters. By comparing the point source distribution in
`blank field' observations with that found in cluster observations,
the number, flux and radial distribution of the sources associated
with the cluster can be determined statistically. In addition the
{\it Chandra} field of view allows AGN to be detected accurately up to 8
arcmin from the centre of the field. This method has been used in the past
to investigate small samples of galaxy clusters, but these contain
significant errors due to field-to-field variance. This study is a
significant advance over previous studies, both in size and
methodology. By analysing 182 galaxy clusters the statistical
variance seen in the smaller studies is significantly reduced, and the
properties of the cluster AGN population can be
identified. Furthermore a sample of this size can be split into
sub-samples and still produce significant results. The dependence of
the AGN population on cluster redshift, mass (estimated from the X-ray
luminosity) and morphology can therefore be found. This analysis
requires careful data reduction and modelling of the sensitivity of
each observation to point sources, which varies across the image. This
 was performed using an automated pipeline developed for this
purpose.

\noindent The key steps in investigating the point sources in the cluster
observations are as follows:

\begin{itemize}
\item{Observations of galaxy clusters with published redshifts $> 0.1$ and `blank'
fields from the {\it Chandra} archive are selected and reduced.}
\item{Each image is visually inspected to insure that the cluster is
detected at the expected location, and that the image does not contain
multiple clusters. The luminosity of each cluster is found and used to
estimate the effects of gravitational lensing on the background
sources. The cluster luminosity and assigned morphological class (see Section
\ref{section_clustermorph}) also allow a later comparison of the AGN content of clusters
as a function of cluster properties.}
\item{Point sources are identified in the fields, and
their properties are calculated.}
\item{For each observation a `flux-limit map' is produced, showing the
detection sensitivity at each point on the image. This accounts for
the detector response, size of the PSF, and the
level of background emission, particularly from the intra-cluster medium.}
\item{The Log N($>S$) -- Log S distribution (where N is the number of
sources and S is the flux) is calculated for each blank and cluster
field, taking into account the sky area sensitive to sources of each
flux value.}
\item{The radial distribution of sources, as a function of distance
from the cluster centre, is calculated. A predicted radial
distribution, assuming no cluster AGN, is produced from the blank
field source distribution and the flux-limit map.}
\item{The effects of gravitational lensing of background X-ray sources
by the galaxy cluster are modelled, and the Log N($>S$) -- Log S
distributions and predicted radial distributions are corrected for
this effect.}
\end{itemize}

The number, flux and radial distribution of the X-ray sources in
clusters can then be determined statistically by comparing the actual
results for each cluster with the prediction, which assumes that no
cluster AGN exist. Section \ref{initialselection} describes the data
reduction and sample selection, for both cluster and blank
fields. Section \ref{Point_sources} explains the source detection, and
Section \ref{Predictions} the model for producing a predicted
distribution. Section \ref{results} explains the first results of this
study. Further results will be published in an accompanying paper.

\section{Initial data reduction and sample selection}\label{initialselection}

There were around 700 imaging observations marked as `Clusters of
Galaxies' in the {\it Chandra} archive in mid-2007.  However the majority of
these are not valid for this study, for a range of reasons.  In order
to determine which observations are useful it is necessary to first
reduce the data, as only then can the reality and position of the
cluster be found. An initial sub-sample of these observations was
therefore put through the first stage of the automated pipeline before
the final sample was defined. This sub-sample contained all
observations with published redshift $> 0.1$ and exposure time $> 10$ ksec. The
details and reasons for these cuts, and the further restrictions
applied to produce the final sample, are described in Section
\ref{thesample}.

\subsection{Data reduction}\label{Data_reduction}
To reduce the initial cluster and blank field samples, an automated
pipeline was developed, using a range of {\sc CIAO} tools and other
programs.  This ensured that the reduction was uniform, and allowed
the whole sample to be reduced efficiently.  In order to obtain the
maximum number of sources around each cluster, all four ACIS-I chips
were used for observations focused on the ACIS-I array, and the three
chips nearest to the aim-point were used for ACIS-S observations (or
less if not all were turned on).  Due to the off-axis degradation of
the PSF the other chips were not investigated as the errors become too
large. Parts of the selected chips were later excluded as the analysis
was restricted to a maximum radius from the aim-point (see Section
\ref{Model}).

For each observation the data were re-reduced from the level 1 event
list using standard {\sc CIAO 3.0.1} tools. The {\it
fix\_batch}\footnote{see
http://cxc.harvard.edu/cal/ASPECT/fix\_offset/fix\_offset.cgi} script
was used to check and correct the astrometry for systematic aspect
offsets.  A time dependent charge transfer inefficiency correction was
applied in all observations taken after 29 January 2000\footnote{after
the ACIS focal plane temperature was lowered.
http://cxc.harvard.edu/cal/Acis/Cal\_prods/tgain/index.html}.  A new
level 2 file was created, using CALDB 2.26 to correct for the
degradation of the QE. The data were filtered for standard grades,
status=0 and the default good time interval (GTI). Further GTI
filtering was performed for each chip by manually masking the
brightest sources and filtering for count rates more than 3-sigma
above the quiescent value. CCD 8 was destreaked using the standard
tools, and the data were filtered for bad pixels. Finally, the data
were filtered for energies between 0.5 and 8 keV to allow better
detection of AGN.

\subsection{The cluster sample and cluster properties}\label{thesample}

An initial sample of observations from the `Clusters of Galaxies'
category, with (probable) redshift $> 0.1$ and exposure time $>$ 10
ksec was selected for this project. Lower redshift clusters were
excluded as only the central regions would be covered by the {\it Chandra}
image.  For example a $z=0.1$ cluster observed with the ACIS-S array
would be observed to a radius of at least 220kpc in all directions,
and with ACIS-I this increases to at least 440kpc. The maximum radius
covered is significantly larger than this, as the cluster is
rarely placed in the centre of the array.

Regardless of the cluster selection criteria the sample will be
heavily biased, as clusters are selected depending on the requirements
of the observer. In particular the sample will be biased towards
relaxed clusters, which are used to constrain cosmological parameters
(e.g. \citealt{Allen}), and rich, highly disturbed clusters, used to
study cluster mergers. Other observations were searches for cluster
emission. By examining the proposal abstracts clusters were
excluded if they were deliberately targeted due to their lensing of
background QSOs. The remaining biases in the sample selection were
parameterised as far as possible by examining the X-ray properties of
the clusters, and taken into account later in the analysis.

The cluster redshift was determined from sources in the NASA
Extragalactic Database (NED).  Observations were selected which have a
confirmed galaxy cluster, cD galaxy, QSO or galaxy overdensity at z $>
0.1$ within $5\arcmin$ of the aim-point.  The archive was examined up
to April 2007 and 192 targets were selected, of which 34 were observed
on more than one occasion (with the same detector array). A small
number of cluster observations fulfilled the above criteria but were
not suitable for the pipeline due to non-standard settings which were
not easily incorporated into the data reduction.

Five properties were evaluated for each cluster field -- the reality,
number of clusters, centre, morphology and luminosity. The first two were
used to reject clusters with no X-ray emission, which are possibly not
true clusters, and fields with multiple clusters at different
redshifts. The X-ray position of the cluster is important as the AGN
distribution may depend on cluster radius, and many optically
discovered clusters have poorly defined centres. The latter two
properties are evaluated in order to determine whether the cluster
properties affect the number or distribution of AGN. The luminosity
also provides an estimate of the mass, which can be used to correct
the predicted source counts for each image for the effect of
gravitational lensing, as described in Section \ref{Lensing}.

\subsubsection{Cluster reality and spatial properties}\label{section_clustermorph}

The morphology, centre and reality of each cluster, and the number of
clusters in each field were determined by examining the 0.5--8 keV
images and the smoothed background images (with point sources removed,
see Section \ref{Model}). The cluster centre was taken to be
the peak of the smoothed background image. 
In the few cases where the cluster consisted of
two peaks of similar brightness, the mid-point was chosen.

\begin{figure}
\begin{center}
\includegraphics{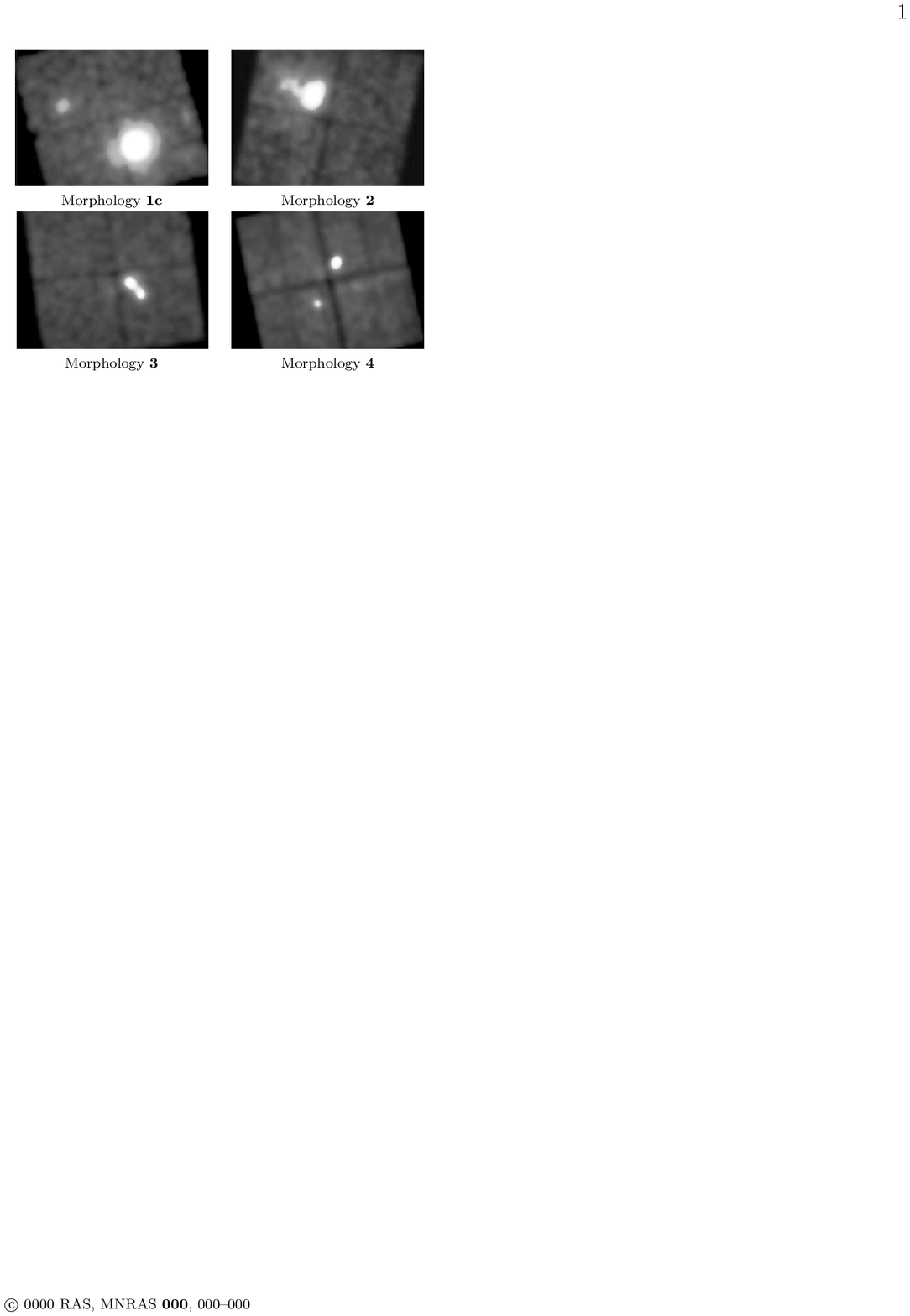}
\caption[Examples of cluster morphologies]{Examples of cluster morphology categories. The images are around 18 arcmin x 12 arcmin.}\label{morphology_fig}
\end{center}
\end{figure}

The reality and morphology were determined by eye, using the full and
smoothed images (described in Section \ref{Model}). The
morphological classifications clearly involve a certain amount of
subjectivity, but they will be sufficient to identify the most
disturbed clusters. The following categories were used, and are
illustrated in Figure \ref{morphology_fig}.

\renewcommand{\labelitemi}{$ $}
\begin{itemize}
\item{{\bf 0.} No cluster emission visible against the background fluctuations.}
\item{{\bf 1}. One relaxed cluster. It may be elliptical or have
edge structure, but not enough to fall into another category.}
\item{{\bf 2}. One disturbed cluster. The disturbance must be such
that the cluster is clearly not simply elliptical or an asymmetric
ellipse, and must be joined to the cluster by visible emission.}
\item{{\bf 3}. Merging cluster. A double-peaked system, with a sub-cluster
with peak emission (in the smoothed image) $> 20$ per cent of the main cluster peak, joined to the main cluster by visible emission or at the same redshift.}
\item{{\bf 4}. Two clusters. A second cluster with peak emission $>
20$ per cent of the main cluster peak but not clearly associated with it.}
\item{{\bf 1c, 2c, 3c.}  As {\bf 1, 2} and {\bf 3}, but with a small
contaminating cluster or group in the field of view. The secondary emission
must have a peak value of $< 20$ per cent of the main cluster peak and not
be clearly associated with it. The contamination can also be an
optically confirmed cluster with no X-ray emission, which is in or
very near the field of view, as described below.}
\end{itemize}
\renewcommand{\labelitemi}{$\bullet$}

\begin{table}
\begin{center}
\begin{tabular}{ccccccccccc}
&&\multicolumn{8}{c}{Observer 1}\\
&&\vline&{\bf 0}&{\bf 1}&{\bf 1c}&{\bf 2}&{\bf 2c}&{\bf 3}&{\bf 3c}&{\bf
4}\\\hline 
\multirow{8}{*}{\rotatebox{90}{Observer 2}}
&{\bf 0} &\vline& 10&&&&&&&\\ 
&{\bf 1} &\vline&&115&4&2&&&&\\ 
&{\bf 1c}&\vline&&&13&&1&&&2\\ 
&{\bf 2} &\vline&&4&&14&&&&1\\ 
&{\bf 2c}&\vline&&&1&&1&&&\\ 
&{\bf 3} &\vline&&&&&&13&&1\\ 
&{\bf 3c}&\vline&&&&&&&&\\ 
&{\bf 4} &\vline&&&&&&&&10\\ 
\end{tabular}
\caption[Morphologies of clusters]{Morphology classes assigned to
$0.1<z<0.9$ clusters by two observers (using X-ray and optical
data).}\label{Cluster_categories}
\end{center}
\end{table}

The images were assigned a category by
two observers, who were broadly in agreement as seen in Table
\ref{Cluster_categories}. The few discrepancies are mainly due to
small contaminating clusters which may be background fluctuations or
faint undetected point sources, and cases of uncertainty over the
degree of disturbance. The morphologies of Observer 1 (the first
author) were adopted as they are slightly more conservative.

The optical data (from the NED) for the cluster fields were used to
check for optically detected clusters in or near to the field of view
(within 15\arcmin from the cluster centre). Three fields were moved
from morphology class {\bf 1} to {\bf 1c} as they contained optically
confirmed clusters at a significantly different redshifts from the
main cluster.  The images were compared to the NED to check that the
detected peak in the X-ray emission corresponded to the location of
the galaxy cluster as given in the NED; clusters were accepted if the
centre of the X-ray emission was within $3\arcmin$ of the NED object.
Two cluster observations were removed as their optical position was $>
3\arcmin$ from the observed X-ray peak, and therefore the optical
redshift may not apply to the X-ray detected cluster.  For
clusters that had more than one redshift measurement in the NED the cluster
redshift was accepted if $\delta z/z < 0.1$. Otherwise the literature
was examined in detail to determine the most accurate redshift -- these
clusters are flagged in Table \ref{Cluster_tables}. In addition one
bright cluster has a slightly revised redshift due to an iron emission
line in the X-ray spectrum, as described in Section \ref{Cluster_lum_temp}.

\subsubsection{The final cluster sample}\label{section_finalsample}

The 192 cluster fields were split into samples depending on their properties.
The final cluster sample contains only uncontaminated confirmed
clusters to ensure that the analysis is not affected by additional
clusters in or near the field of view, which could also contain AGN
and may contribute to the lensing of background AGN (see Section
\ref{Lensing}, although this is likely to be minimal in most cases).
The final sample therefore consists of the 148 observations with morphology
class {\bf 1, 2} or {\bf 3} and $0.1<z<0.9$.

The twenty weakly contaminated cluster fields ({\bf 1c, 2c} and {\bf
3c}) were included in a second sample, as the fields may still be of
interest. The ten fields which clearly contained a second cluster
(type {\bf 4}) were rejected from the rest of the analysis.

The fourteen $z>0.9$ cluster observations with secure redshifts were placed in
a third sample, regardless of the reality or extent of their emission,
as at this redshift range almost all of the observed clusters are centred on
active galaxies, and often the extended emission may be too faint to
detect or be contaminated by AGN jets  Some of these
objects are better classified as proto-clusters, so they are analysed
separately from the rest of the sample. 

The final cluster fields, split into the above categories, are described in
Table \ref{Cluster_tables}.

\subsubsection{Cluster luminosities and temperatures}\label{Cluster_lum_temp}
To compare clusters at different redshifts, luminosities need to be found
in the same rest-frame band for each cluster.
A spectrum was extracted from the level 2 data for each cluster and
fit with a thermal model, which was then evaluated in the given
band. The following analysis was applied to all cluster observations,
but is only truly valid for clusters with morphology classes 1--3 as
listed in Table \ref{Cluster_tables}. 

Spectra were extracted from the 0.5--8 keV band data (to simplify the
data reduction) from circular apertures centred on the cluster
centre. The chosen aperture included $\sim 99.5$ per cent of the cluster
counts, and the background spectrum was taken from an annulus with
radii of 1.1 and 1.49 times the cluster radius. Point source regions
were subtracted and the regions containing the brightest point sources
were enlarged if necessary, to ensure that they did not contaminate
the cluster emission. Areas of bad or no exposure were also removed.
Response functions were calculated for the central region of the
cluster aperture, rather than finding a weighted response over the
full aperture, due to the time required for the latter.  Tests on
three clusters found that the difference in flux for a single central
response file compared to that for the full region was $< 2$ per cent, which
is negligible compared to the errors in the model.

Spectra were fitted using an absorbed Raymond--Smith model
\citep{RaymondSmith} in {\sc XSPEC v11.3.1}, binned to a minimum of 25
counts per energy interval. The galactic neutral hydrogen density was
fixed at the local values \citep{Dickeylockman}, and the redshift
fixed to the value in Table \ref{Cluster_tables}. For clusters with
more than one observation the multiple spectra were fit
simultaneously. The model errors are underestimated as they do not
take into account errors due to taking the calibration of the central
pixel only. To get a better measure of the accuracy of the
luminosities, the spectra of clusters which were observed twice were
fitted individually, and the difference between the luminosities were
found to be less than 0.1 dex at all fluxes. The observed luminosities
of the clusters using this method are given in Table
\ref{Cluster_tables}. These generally match the {\it Chandra} luminosities
in the literature (e.g. \citealt{NEDwg}) to within 10 per cent.

\subsection{Blank fields}\label{blank_fields_sec}

It is necessary to have a control sample of blank fields in order to
calculate the expected distribution of point sources in each cluster
observation due to foreground and background objects (which will be
referred to as `background sources', although they may in fact be in
the foreground).
To avoid biases due to large scale structure and
statistical variance due to low counts it is desirable to have as
large a sample as possible of blank fields.

Many X-ray surveys of `blank' fields have been conducted in order to
study the general X-ray source population.  Some of these observations
were selected from the archive, and reduced with the pipeline to
ensure consistent data reduction. Fields that contained galaxy
clusters which were discovered independently of the blank field
observation were removed, but fields with serendipitous cluster
detections were retained.  Individual pointings were selected so as to
maximise the sky area and match the depth of the cluster observations.

In addition to the `true' blank fields, observations of high-redshift
($z>2$) quasars or radio galaxies were also used. These were added to
increase the sample size, and hence reduce the errors due to low
source counts (particularly at high fluxes). In addition, all of the
blank field observations used the ACIS-I detector, so observations
using ACIS-S were required to test for differences due to the
detector. The fields all have observation times $>10$ksec and
redshifts in the NED.  In most of these fields the QSO is visible at
the aim-point and it is possible that there are extra sources at the
redshift of the QSO due to either clustering or lensing (these will be
rare as the observations are shallow and the target QSOs very
distant).  In all images a circle of radius $25\arcsec$ was removed
from around the aim point, as this radius excludes all other objects
identified in NED at the QSO redshifts (with the exception of one
field, which was rejected). These regions were excluded from the
analysis using the masks described in Section \ref{Model}. One field
had a further region removed due to a rare serendipitous detection of
a nearby galaxy with resolved point sources.

Once the data were reduced, the blank field number counts were checked
to ensure that including the high redshift QSO fields does not bias
the background (as explained in the Appendix).  The final sample of
blank fields consists of 22 true blank fields and 22 QSO fields, which
are listed in Table \ref{blankfieldtable}.

\section{Point source detection and properties}\label{Point_sources}

\subsection{Source detection}\label{Source_detection}
Images were made using unbinned data and exposure maps (in
s$^{-1}$cm$^{-2}$) were made assuming the sources have a photon index
of $\Gamma = 1.7$, typical of unobscured AGN at the sample flux limits
(see for example figure 3 of \citeauthor{ChandraDFS}, 2001). Tests on
a few images showed that changing the spectral index to other
realistic values does not significantly change the sources detected or
their significances. Sources were detected using the {\sc wavdetect}
package \citep{Wavdetect}, with wavelet scales of 1,2,4,8 and 16
pixels and a significance threshold of $10^{-6}$. Tests using
different wavelet scales suggest that $ \ll 1$ per cent of sources are close
enough to be missed by using these scales, but would be detected using
scales separated by $\sqrt{2}$.  The source list output from {\sc
wavdetect} was examined by eye to remove detections of the extended
cluster emission. A montecarlo simulation of the source detection
(Appendix A.1) shows that very few sources are missed by
{\sc wavdetect}, even accounting for the rapidly varying background in
regions near the cluster centres.

Many clusters and blank fields were observed more than once, and where
possible in these cases data from up to three observations were merged
before the sources were detected, to give far deeper images and
maximise the number of sources. Images were only merged if the same
detector (ACIS-I or ACIS-S) was used. The process is similar to that
for single observations with the following additions:--

The astrometry of the images was adjusted using the {\it align\_evt}
routine\footnote{ALIGN\_EVT v1.6, written by Tom Aldcroft} as, even
after correcting the aspect files, small offsets often exist between
images. Images were matched using sources detected in the central 4
arcmin of each image, where the PSF is smallest.  
Individual images were made, and exposure maps created for each
observation. A combined image and combined exposure map were computed.

{\sc wavdetect} determines whether a source is real based on the
source extent and the size of the PSF, which is complex for merged
images with different aim points. In this case the combined PSF size
at each point was calculated by combining the PSF sizes of the
individual images, weighted by their exposure map, and calculating the
3-sigma encircled energy size of the resulting source. 
These PSF sizes were input into
{\sc wrecon}\footnote{Using a more flexible version, kindly provided
by Peter Freeman (private communication).} in order to give detections
and sizes that are comparable to the standard {\sc wavdetect} results
for single images.

\begin{figure}
\begin{center}
\includegraphics[scale=0.4,angle=0,clip]{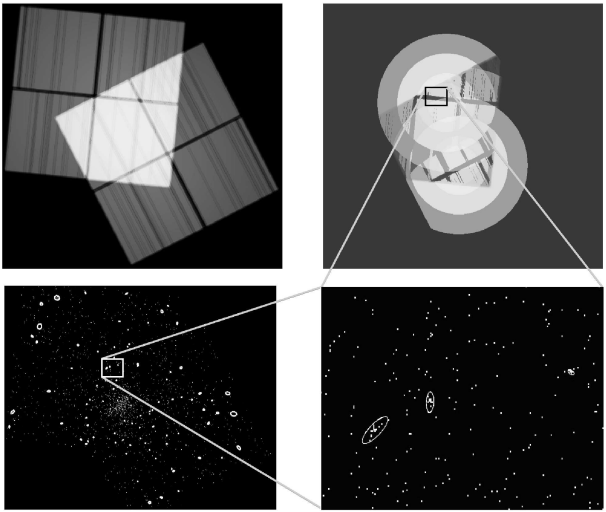}
\caption[Source detection inputs and outputs for MACS J1149+22]{Source
detection inputs and outputs for MACS J1149+22. The top panel shows
the exposure map (left, higher exposure is white) and calculated
expected PSF size distribution (right, smaller PSF is whiter) for this
observation. 
The lower panel shows the sources detected on the
combined image, using these inputs, and a enlarged portion of the
image, covering an area where the expected PSF size
varies rapidly. The full image is $\sim$24 arcmin across, and the pixel size is $\sim$0.5 arcsec. 
 It is clear that the input PSF size
distribution, combined with the detection power of {\sc wavdetect},
accurately finds the true source centre and extent. Some detected
sources could be background fluctuations, but these are later removed
as described in Section \ref{Point_sources}.}\label{mj1149_eg}
\end{center}
\end{figure}

As an illustration of this technique, Figure \ref{mj1149_eg} shows the
sources detected in a combined image of MACS J1149+22. The combined
exposure map and expected PSF distribution are also shown.

\subsection{Source properties}\label{Source_counts}
It is important to determine the source properties very accurately, as
only a few percent of the detected sources are likely to be cluster
sources. The observations have a wide range of exposure times, and the
source sizes also change with off-axis angle, and small errors in the
determination of the source properties could therefore wipe out any
signal from the cluster sources, or introduce biases with, for
example, redshift or exposure time.
 
Because of the need for high accuracy the reality and properties of
the {\sc wavdetect} detected sources were re-determined using more
stringent criteria. This is also necessary in order to create an
accurate model of detection probability, taking into account the
effect of the variable background in cluster fields, as described in
Section \ref{Predictions}. {\sc wavdetect} outputs were used to
determine source positions and sizes, but other properties of the
sources, such as counts and significance, were re-calculated. The
positions and properties of some of the significant sources detected
in one field are listed in Table \ref{sourcetable}, which also
contains the web address of the full source list for all cluster
fields.

In order to maximise the signal-to-noise for individual sources, the
{\sc wavdetect} source sizes were used to determine a circular
aperture size for each source (as the current {\it Chandra} PSF models are
only measured at a few radii).  The aperture had radius $1.2 \times
r_{\rm max}$, where $r_{\rm max}$ is the semi-major axis of the {\sc
wavdetect} 3-sigma ellipse. Extensive testing showed that this radius
of aperture maximised the signal-to-noise for the sample whilst
minimising the missed source counts.

In deep X-ray images many sources either overlap with other sources or
with areas of bad exposure such as chip gaps.  Pixels in the source
aperture were rejected if they were within the aperture of another
source, or had exposure below 10 percent of the median value for the
source. These pixels were replaced by their reflection on the opposite
side of the aperture if possible, or otherwise with other pixels from
the same radius. Around 3 per cent of the sources detected required some
degree of correction, and for $< 0.5$ per cent of sources the correction is
only accurate to within a factor of $\sim$2 due to the large
correction area.

For each source the mean background count rate per pixel was
calculated in an annulus of area 10000 pixels, with an inner radius of
$1.5 \times r_{\rm max}$.  Any pixels in this area within $1.5 \times
r_{\rm max,i}$ of a nearby source, i, or with exposure less than 10
percent of the median, were rejected.  Because of the large variations
in the PSF, this method works far better than an annulus scaled with
aperture size and the effect of highly varying background, such as
around clusters, on the source flux was found to be negligible.

The source counts are given by
\begin{eqnarray}
	{\rm Counts} = C_A - \rm{Bkg}\label{eq1a}\\
	{\rm Bkg}= C_B \frac{N_A}{N_B} \frac{E_A}{E_B}\label{eq1b}
\end{eqnarray}
where $C$ is the total counts in a region. $E$ is the mean exposure
map value of good pixels, and $N$ the number of good
pixels. Subscripts $A$ and $B$ refer to the source aperture and
background region respectively. $C_B$ is scaled by the ratio of the
exposures as in $\sim 18$ per cent of sources the mean background and
aperture exposure differ by over 10 percent.

Throughout the
calculations the Gehrels \citep{Gehrels} approximation $G(C) = 1+\sqrt{0.75+C}$
is used to approximate both the Poissonian 1-sigma upper and lower limit.
			   
\noindent The error on the counts is given by
\begin{equation}\label{eq2}
	\sigma_{\rm Counts}^2 =
	\left(G(C_A)\right)^2+\left(\sigma_{\rm Bkg}\right)^2
\end{equation}
where the error on the calculated background counts in the annulus,
$\sigma_{\rm Bkg}$, is
\begin{equation}\label{eq3}
	\sigma_{\rm Bkg}^2 = \left( G(C_B)
	\frac{N_A}{N_B} \frac{E_A}{E_B}\right)^2 + \left( G\left( C_B
	\frac{N_A}{N_B} \frac{E_A}{E_B} \right)\right)^2
\end{equation}
which is the combination of the error on the estimation of the background
count rate, and the error on applying this (low) background value to the
aperture.

\noindent The source signal-to-noise ratio (SNR) is then 
\begin{equation}\label{SNR}
SNR = {\rm Counts}/\sigma_{\rm Counts}
\end{equation}
and the significance, SIG (following \cite{Olivia}) is defined as   
\begin{equation}
C_A = {\rm Bkg} + {\rm SIG}\times \sigma_{\rm Bkg} \\
\end{equation}
so that
\begin{equation}
{\rm SIG} = {\rm Counts}/\sigma_{\rm Bkg}.\label{SIG}
\end{equation}

A cut of $SIG > 3$ was applied to construct a catalogue of real sources.  A
significance of above 3 means that the source is not a background
fluctuation with above a 3-sigma probability. The correlation between $SIG$
and the SNR is very good, with a significance cut of 3 corresponding to a
SNR of around 1.5.  This cut is more conservative than the {\sc wavdetect}
significance parameter, and produces a more robust source list. On average
it reduces the {\sc wavdetect} source list by around 18 per cent.

To calculate fluxes (in erg cm$^{-2}$ s$^{-1}$, for the 0.5-8 keV band) the exposure map value at each pixel (in cm$^2$ s) was combined with the counts;
\begin{equation}\label{eq4}
	{\rm Flux} = \frac{1}{k} \times \left( \sum_{i\in A} \frac{C_i}{E_i} -
	\frac{N_A}{N_B}
	\frac{E_A}{E_B} \sum_{i\in B} \frac{C_i}{E_i}  \right)
\end{equation}
where the summation over $i$ is over the individual pixels in a
region. $k$ is the conversion from counts to ergs assuming the source
has a spectrum with $F_{\nu} \propto \nu^{-1.7}$ between 0.5 and 8
keV, and energy dependent absorption by galactic hydrogen following
\cite{MorrisonMcCammon} with column density from \cite{Dickeylockman}.

The flux missed by choosing a smaller aperture is $\sim 1$ per cent for the
brightest sources and $\sim 4$ per cent for the faintest, depending on the
source counts. This small correction factor was applied to the source
fluxes to eliminate errors in the full population caused by
differences in exposure times. Luminosities were calculated in the 0.5
-- 8 keV emission band assuming a $F_{\nu}\propto \nu^{-1.7}$ spectrum.

\section{Predicted source distributions}\label{Predictions}

To interpret the number counts of point sources in each image an
accurate model of each observation is required to determine the number
of sources expected if there were no AGN in the cluster. This model
requires the minimum flux detectable at each pixel and the number of
blank field sources as a function of flux. The changes in sensitivity
are particularly important in the cluster fields as the extended
cluster emission may obscure faint central sources. The minimum flux
model is described below. Section \ref{Lognlogssection} describes the
calculation of the expected number of sources for each observation,
and Section \ref{Lensing} explains the correction of this prediction
due to gravitational lensing by the cluster.

\subsection{Modelling the sensitivity of each observation}\label{Model}
A flux limit map was computed following the method of
\cite{Olivia}. From equation \ref{SIG}, the counts for a source
centred at pixel i and detected at the minimum significance of 3 is
$C_{min, i} = 3 \sigma_{\rm Bkg,i}$, which combined with equation
\ref{eq3} and the conversion to flux used in equation \ref{eq4} gives
a minimum flux detectable with significance $>3$ at pixel i of

\begin{eqnarray}
	S_{\rm min,i}\hspace{-0.3cm} &=&\hspace{-0.3cm} \frac{C_{\rm min,i}}{E_i k}\\
\hspace{-0.3cm} &=& \hspace{-0.3cm}\frac{3}{E_i  k} \left( \hspace{-0.1cm}  \left( G\left( R_{B,i} N_{B,i}\right) 
	\frac{N_{A,i}}{N_{B,i}}\right)^2 \hspace{-0.2cm}+ \left( G(R_{B,i}N_{A,i})
	\right)^2  \hspace{-0.1cm}\right)^{1/2} \label{Smin}
\end{eqnarray}

\noindent where $S_{\rm min,i}$ is the minimum flux detectable with significance $>
3$ at pixel i. Subscript A indicates values for the predicted source and B
the predicted background, and R is the rate in counts pixel$^{-1}$sec$^{-1}$. 
The inputs for the prediction are then the exposure $E_i$, source size
$N_{A,i}$ and background count rate $R_{B,i}$ at each point on the
image.

The exposure is simply the sum of the individual exposure maps
described in Section \ref{Source_detection}. There are regions where
the gradient in exposure will make source detection difficult, and
these are masked out later as described below. The
errors in the exposure map should be small and are not easily
calculable. As they will affect both the blank fields and the cluster
fields in the same way they can be neglected here.

\begin{figure}
\begin{center}
\includegraphics[scale=0.8,clip]{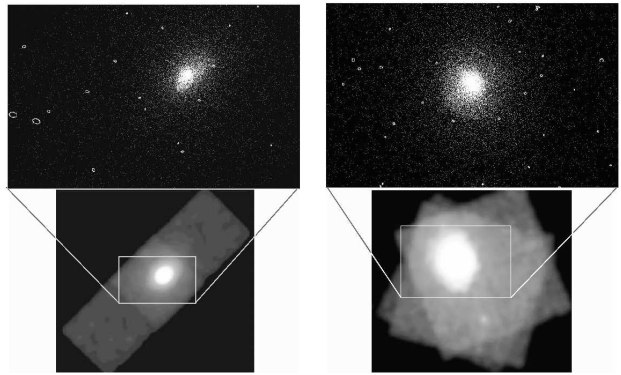}
\caption[Background images for two example fields]{Background images for two
cluster observations, Abell 1068 (left) and RX J1720+26 (right). The top
panel shows the the central region of the images, with sources removed and the source regions
marked by ellipses. The bottom panel shows the full smoothed background
images (with a square root scale).}\label{Background_eg}  
\end{center}
\end{figure}

For each observation the background rate, including the diffuse
cluster emission, was calculated by replacing the point sources with
local background and smoothing the image with a gaussian kernel of
radius 40 pixels. Figure \ref{Background_eg} shows an example of the
background images produced.  To find the error on the background it is
easiest to assume that the smoothed background rate is given by the
average of the counts in a circle, rather than calculating the errors
on the true gaussian convolved image. In other words, $R_{B,i} \approx
\sum_{j\in {\rm Area}}C_j/{\rm Area}$ where the area is a circle of
radius 40 pixels centred on i. This gives a simple equation for the
error -- $\sigma_{R_{B,i}}\approx \sqrt{\frac{R_{B,i}}{\pi 40^2 }}$.
To test the model background, the background rate for each detected
source (using aperture photometry) was compared to that from the
smoothed images at the same position. The model background accurately
reproduces the calculated background for the detected sources (with
SIG $>3$), with no systematic offset. 

\begin{figure}
\begin{center}
\includegraphics[scale=0.45,angle=0,clip]{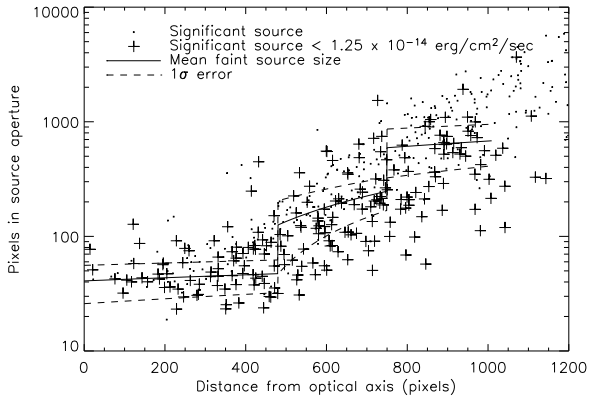}
\caption[Actual and model source size distribution]{Source size as a
function of distance from the optical axis (four representitive blank
fields are shown to avoid overcrowding). The mean size and 1$\sigma$
errors ware calculated for the faint sources. This is then the expected
source size for sources near the flux limit at each
pixel.}\label{Model_psf}
\end{center}
\end{figure}

The expected source size distribution was calculated using the
apertures for the detected sources from 8 representitive blank fields,
and checked against the detected source sizes in all fields. Apertures
derived from the {\sc wavdetect} output were used instead of the given
PSF size as this is how the source properties were determined.  The
radial distribution of aperture sizes is shown in Figure
\ref{Model_psf}, which also shows the chosen model radial source size
distribution. This model was determined from the data for significant,
low flux sources ($S < 0.25 \times 10^{-14}$ erg cm$^{-2}$s$^{-1}$)
which are at the detection limit of these observations.  The aperture
size was found to jump at radii of 480, 750 and 1010 pixels, due to
the behaviour of the PSF combined with the wavelet scales chosen (this
is illustrated by the fact that the brighter sources, marked by dots
in Figure \ref{Model_psf}, are far closer to a constant slope, as they
are affected less by the wavelet scales and trace the true change in
PSF). The aperture size was modelled between each jump with a best-fit
quadratic, and the 1-sigma error was determined by the distribution of
sources around this fit.  Above a radius of 1010 pixels the aperture
size jumps considerably so this area was removed from the calculation.
 
Again, comparison with the actual source sizes shows that this model
is accurate to within the errors and has no systematic error. There
were no significant differences between source sizes in the ACIS-I and
ACIS-S chips.  For multiple observations the source size distribution
was calculated for each observation, then combined weighted by
exposure map.

A mask was constructed to restrict the area to regions where the model
is accurate. This removes the effects of chip gaps, chip edges and
errors in the modelling. Edge effects, especially due to the
background smoothing, affect areas within 60 pixels of the image edge,
and 40 pixels of chip gaps, so these areas were removed.  For merged
images, $i$, the `chip boundary' area was included if $ E({\rm
max})_{i} < 0.5 \times \Sigma_{j=good} E({\rm max})_{j}$, where
$E$(max) is the on-axis exposure of an image and the sum is over all
images, $j$, with good exposure in the `chip boundary' area.  As
described in Section \ref{Model}, regions where the model
source size is greater than 700 pixels were also removed.

\begin{figure}
\begin{center}
\includegraphics[width=\columnwidth,clip]{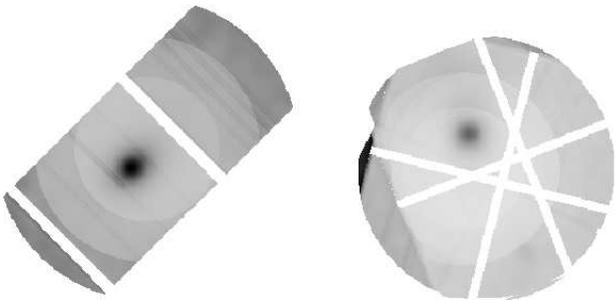}
\caption[Final flux limit model for two observations]{The final flux
limit model for Abell 1068 (left) and RX J1720+26 (right). Light areas
are the most sensitive and have the lowest limiting flux (1.7 and 1.2
$\times 10^{-15}$erg cm$^{-2}$ s$^{-1}$ respectively), and dark grey
are the least sensitive (7.5 and 1.1 $\times 10^{-14}$erg cm$^{-2}$
s$^{-1}$). The PSF size, exposure map and cluster background all
clearly affect the final limiting flux distribution.  }\label{Slim_eg}
\end{center}
\end{figure}

The final flux limit model for the two example fields is shown in
Figure \ref{Slim_eg}, where all cuts and masks have been applied.  To
check the flux limit model the fluxes of all detected sources were
compared to the minimum flux detectable at the source position. Almost
all ($>97$ per cent) of sources are brighter than the flux limit at their
position. Those that are slightly fainter than the corresponding flux
limit have large errors on their flux, so that $\ll 1$ per cent of sources
are over $1 \sigma$ fainter than the calculated flux limit at their
position.

\begin{figure}
\begin{center}
\includegraphics[trim=0.5cm 0cm 0cm 0cm,clip,width=\columnwidth]{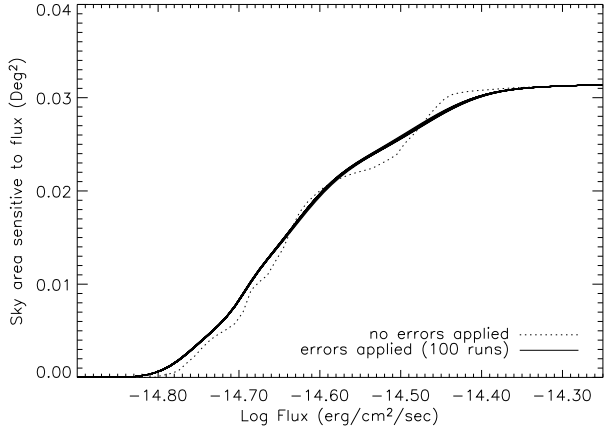}
\caption[Errors on the sky area sensitive to $S$]{The sky area
sensitive to sources of flux $>S$ for one field. 
when no errors are applied to $S_{\rm min}$ is shown by the dotted
line and 100 examples of when random errors are applied by the solid
lines. The variation in the solid lines is negligible compared to the error
on the number of sources.}\label{Area_error}
\end{center}
\end{figure}

The combined effect of the errors on $S_{{\rm min},i}$, summed over
the image, is not straightforward to calculate. Random errors were
added to the calculation for each pixel, and the sky area at each flux
re-calculated. As the errors on the background level are correlated
between pixels, the error in each 80 $\times$ 80 pixel square were
changed by the same (randomly selected) number of sigma. Changing the
size of this region did not change the results. The error in
calculating the flux conversion factor, k, was not included as this
will affect each field in the same way.  Figure \ref{Area_error} shows
the effect of these errors on one field. The sky area without errors
has quite steep jumps due to the sudden changes in the model PSF (due
to the wavelet scales), but applying random errors to the flux limit
at each pixel smooths this distribution.

\subsection{Log N($>$S) -- Log S and radial distributions}\label{Lognlogssection}

The number of sources brighter than a given flux is calculated for each
field, or for a combination of fields, to produce a plot of Log N($>$S)
against Log S, using
\begin{equation}
{\rm Log N}(>S_0) = \sum_{S>S_0} \frac{i_S}{A_S}
\end{equation}

\noindent where N($>$S$_0$) is the number of sources brighter than S$_0$, i
is the number of sources of flux S, and A the total sky area available to
detect a source of flux S. 

The errors are dominated by the number of sources detected, so once
the errors have been added to individual pixels (Figure
\ref{Area_error}), the error on the sky area can be neglected. The
 error on the total number of sources was used for the
brightest sources, where the sky area is constant, such that;
\begin{equation}
\sigma_{{\rm N}(>S_0)} = \frac{\sigma(i_{S>S_0})}{A}
\end{equation}
\noindent for $A > 0.99 \times A_{max}$. When the sky area starts to
decrease (at lower flux), $\sqrt{i}$ errors are used as they are
able to take account of the weighting by sky area; at these fluxes the
number of sources is relatively large ($i \gtrsim$10) and the difference
between Gehrels and $\sqrt{i}$ approximations becomes minimal. The error is
then given by
\begin{equation}
\sigma_{{\rm N}(>S_0}) = \sqrt{\sum_{S>S_0} \frac{i_S}{A_S^2}}
\end{equation}

It is worth noting here the effect of the Eddington bias
\citep{Eddington}, whereby random flux errors can increase the
measured source counts above a chosen flux level.
\cite{Mannersthesis} show that the net effect for one field is $\sim
1$ per cent extra sources above $1.1 \times 10^{-15}$ erg cm$^{-2}$ sec$^{-1}$. Since
this is a small effect and will affect the cluster and blank field
samples in the same way, it is not accounted for in the analysis.


For each field, in addition to the Log N($>$S) -- Log S distribution,
the radial distribution of sources was found and compared to the
radial prediction assuming no cluster sources. This was calculated
using the blank field Log N($>$S) -- Log S and the $S_{\rm min}$ map.
Errors on the predicted radial distribution were found by applying the
Log N($>$S) -- Log S distribution with 1$\sigma$ errors to the $S_{\rm
min}$ map.  The predicted and actual radial distributions were
calculated from the X-ray cluster centre, or from the aim-point if no
cluster was visible. 

As a check of the pipeline method the radial and Log N($>$S) -- Log S
distributions were calculated for the 44 blank fields, as described in
Appendix A.3. The pipeline prediction well reproduces
the actual distribution of sources in the blank fields. In addition, checks were made for differences between the two ACIS detectors on {\it Chandra}, as described in Appendix A.4

\subsection{Corrections for gravitational lensing}\label{Lensing}

The effect of gravitational lensing of X-ray sources by the galaxy
cluster is small, but is expected to be significant over many
fields. As discussed by \cite{Refregier}, after lensing the flux of
each source is increased by a factor $\mu_{\theta}$, where $\theta$ is
the angular distance from the cluster centre, and the number density
is decreased by the same factor due to a decrease in the apparent sky
area of the image. Whether this results in a net increase or decrease
in sources at a given flux depends on the slope of the Log N($>S$) -
Log S distribution. In moderately deep cluster observations the slope
of the number counts is shallow, resulting in a deficit of sources in a
lensed field compared to a blank field.

\cite{Olivia} estimate an expected deficit of X-ray sources
of $\sim 10$ per cent in the central 0.5 Mpc of MS1054-0321 (z=0.83). This is
insignificant for a single field but the cumulative effect over many
fields may affect the sample.  In addition, as the effect of lensing
on the number counts is more significant for bright, moderate redshift
clusters, gravitational lensing could bias the results.

To calculate exactly the difference between the cluster and blank
fields that is due to gravitational lensing requires detailed
knowledge of the dark matter distribution in the cluster. As this
study is investigating a statistical excess of sources in a large
number of fields, exact determination of the lensing is unnecessary
(and unfeasible). Instead, the radial loss or gain of sources in each
image due to the cluster is estimated using a simplified model of
gravitational lensing, with the only inputs being the X-ray
luminosity, position in the image and redshift of the cluster, an
assumed background distribution of X-ray sources and the sensitivity
distribution of the observation.

For an NFW mass profile (Navarro, Frenk \& White, 1997\nocite{NFW})
$\mu_{\theta}$ is only dependent on a characteristic radius and the
cluster mass, as shown in Appendix A of \cite{Myers}, using formulae
and data from \cite{Maoz}, \cite{Bartelmann} and
\cite{NFW}. \cite{Maoz} also show that the characteristic radius can
be approximated by a function of the cluster mass. This in turn can be
estimated by using the redshift dependent cluster mass - luminosity
relation in equation 15 of \cite{Maughan}. The cluster X-ray
luminosities and redshifts (see Sections \ref{section_clustermorph}
and \ref{Cluster_lum_temp}) can therefore be used to calculate the
distribution of $\mu_{\theta}$ for each field. Although this
calculation relies on a number of empirical relations, this will not
introduce large errors as discussed below.

Three models of the X-ray background are used, as described
below. They are all calculated for rest-frame 2--8 keV (hard band)
sources which, where the lensing from clusters will be strongest,
corresponds to observed 1--4 keV sources. The lensing factor calculated
in this section is fractional, so only the shape and relative
normalisation matter. It is therefore assumed that the population of
hard sources in the model shows the same distribution and redshift
evolution as the sources in the cluster image.

Two of the models use the \cite{Barger} X-ray luminosity function from
$0.1<z<1.2$, but extend it to z=5. The first reduces the density by a
factor of $z^3$ at high redshifts, which is a reasonable fit to the
sources with confirmed redshifts and is therefore a lower limit. The
second model scales the space density of sources such that the energy
density per comoving volume remains flat at $z>1$. This is the maximum
value allowed by the \cite{Barger} data, so is an upper limit.  The
third model adopted here is a luminosity dependent density evolution
(LDDE) model, with best fit parameters from \cite{Ueda}, which is the
best fit to the hard X-ray luminosity function from the ChaMP survey
(see \cite{Green} and \cite{Silverman2} for details).  The three model
luminosity functions are calculated from z=0 to z=5, in redshift steps
of 0.1. The lower end is important as a lot of sources will not be
lensed, and these will reduce any fractional deficit due to lensing.
The luminosity functions at each redshift are re-normalised to
represent the sky volume visible in an image of 1 deg$^2$, rather than
per cubic Mpc.


The effect of lensing on the model background source distribution is
calculated as a function of cluster-centric distance, cluster redshift
and cluster X-ray luminosity for each field. The lensed luminosity
functions (boosted luminosities and lower space densities) of the
non-cluster sources were found for redshifts 0 - 5 in steps of 0.1 and
were converted to flux distributions in the observed band and summed
over all redshifts. The resulting lensed Log N($<S$) -- Log S
distributions were compared to the unlensed distribution, and the flux
limit at each point in the image, to calculate the fractional change
in sources detected at each pixel. This correction was applied up to
300$\arcsec$ from the cluster centre, and gives a maximum correction
per field of $\sim 1$ source.

The three models for the X-ray background distribution did not give
significantly different results (far smaller than the errors on the
source distribution) so only the LDDE model was used, which gives
results between the two extreme models taken from the \cite{Barger}
data. The largest source of error in this model is if there is a
systematic miscalculation of the cluster properties, but this is still
a small source of error overall. For example if the cluster mass is
assumed to be systematically out by 30 per cent for all clusters then
this would add the equivalent of 1.5 percent to the error bar at 1
Mpc. Random errors in the cluster properties due to scatter in the
cluster scaling relations will generally cancel out over a large sample.

When the correction for gravitational lensing is applied, the total
number of non-cluster sources predicted in an average field
decreases. The prediction for the central 3 Mpc of the 148 good
cluster fields is found to decrease by 0.7 per cent, which is around
0.27 sources per field on average.  The calculated number of sources
in the cluster, which is the number of detected sources minus the
prediction, therefore increases by the same number. This is shown in
Section \ref{Radial_results}, where the excess sources per cluster
field before and after the lensing correction are compared. The number
of predicted sources brighter than $10^{-14}$ erg cm$^{-2}$sec$^{-1}$
decreases by 0.6 per cent, which is 0.06 sources per field. The
lensing correction is small, typically $<0.5\sigma$, but it is not
insignificant as the correction is predominantly in the central
regions. All statistics and plots presented in the
remainder of this paper use the lensing correction, but none of the
results are significantly altered if the lensing correction is
ignored.

\section{Results}\label{results}

\subsection{Excess sources in cluster fields}\label{Excess_results}
\begin{figure}
\begin{center}
\includegraphics[width=\columnwidth,clip]{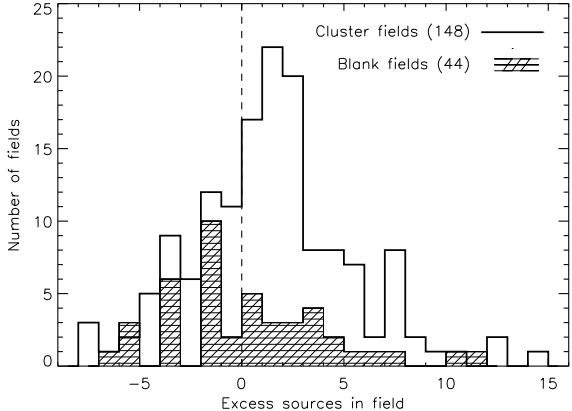}
\caption{The excess number of sources per field, compared to the
prediction, for 44 blank fields and 148 uncontaminated cluster fields
with $0.1<z<0.9$. The mean for the cluster fields is $1.46 \pm 0.32$, compared to $0.47 \pm 0.61$ for the blank fields.}\label{hist1mpc}
\end{center}
\end{figure}

The number of cluster sources in each field was estimated by
subtracting the predicted number of sources (Section
\ref{Predictions}) from the actual number of well-detected sources
(Section \ref{Point_sources}), to get the excess sources in each
field. The results presented in Figure \ref{hist1mpc} show the excess
sources within 1 Mpc of each cluster centre, which is the maximum
radius observed for the lowest redshift clusters. The resulting
histogram shows that cluster fields have a wide spread of calculated
excess sources, including negative values, but that the average excess
is clearly non-zero. For the blank fields, with assigned redshifts
randomly chosen from the redshift distribution of cluster fields, the
excess sources within 1 Mpc of each field are consistent with zero.
Figure \ref{hist1mpc} shows that the galaxy clusters have, on
average, around 1.5 sources each within a projected distance of 1 Mpc.
This value is an average over clusters of different redshifts and
luminosities, and observations of different exposure time; the
dependence of number of cluster X-ray sources on these variables will
be analysed in the next paper in this series.

\begin{figure}
\begin{center}
\includegraphics[width=\columnwidth,clip]{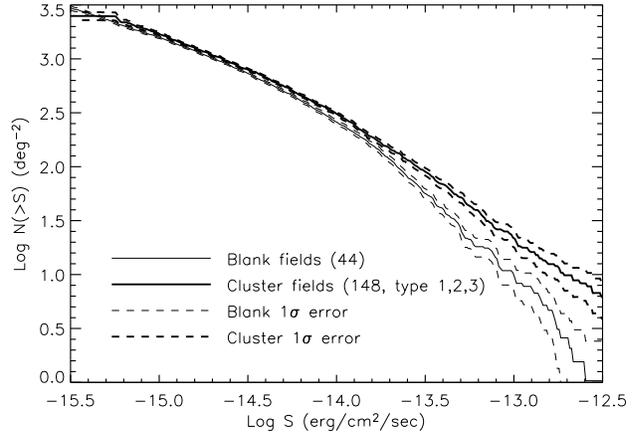}
\caption{The combined Log N($>S$) -- Log S distribution within 1 Mpc of
the cluster centres for the 44 blank fields and 148 uncontaminated
cluster fields with $0.1<z<0.9$. }\label{lognlogs}
\end{center}
\end{figure}

The Log N($>S$) -- Log S distribution was plotted for the blank fields
and the 148 uncontaminated cluster fields with $0.1<z<0.9$. Figure
\ref{lognlogs} shows that the cluster fields have a $\sim2\sigma$
excess at fluxes of $> 10^{-13.7}$ erg cm$^{-2}$sec$^{-1}$. An excess of
$\sim1\sigma$ is seen at fainter fluxes. These are not strongly
dependent on the lensing correction, which changes the results by a
maximum of $0.2\sigma$.  A Kolmogorov--Smirnov (K--S) test on the
sources brighter than $> 10^{-13.7}$ erg cm$^{-2}$sec$^{-1}$ shows that the
cluster and blank field populations differ at the 96 per cent level.

The most significant excess in the Log N($>S$) -- Log S distribution is
found at bright fluxes, but this is partly due to the lower number of
bright background sources. In fact only half of the excess sources
within 1 Mpc have flux $> 10^{-14}$ erg cm$^{-2}$sec$^{-1}$ (a mean of 0.76
$\pm$0.18 per field). Sources brighter than this flux are quite likely
to be AGN, as this corresponds to a (k-corrected) luminosity of $>2.5
\times 10^{41}$ erg sec$^{-1}$ in all clusters, and $>10^{42}$ erg sec$^{-1}$ in over
80 per cent of the sample. In clusters with $z<0.2$ sources with luminosity
$<10^{41}$ erg sec$^{-1}$ can be detected, which are far less likely to be
AGN, but the majority of sources have either flux $\gg 10^{-14}$ erg cm$^{-2}$sec$^{-1}$ or are in higher redshift clusters and so are
likely to be AGN.

\subsection{Radial distribution of cluster sources}\label{Radial_results}

\begin{figure*}
\begin{center}
\includegraphics[width=2\columnwidth,clip]{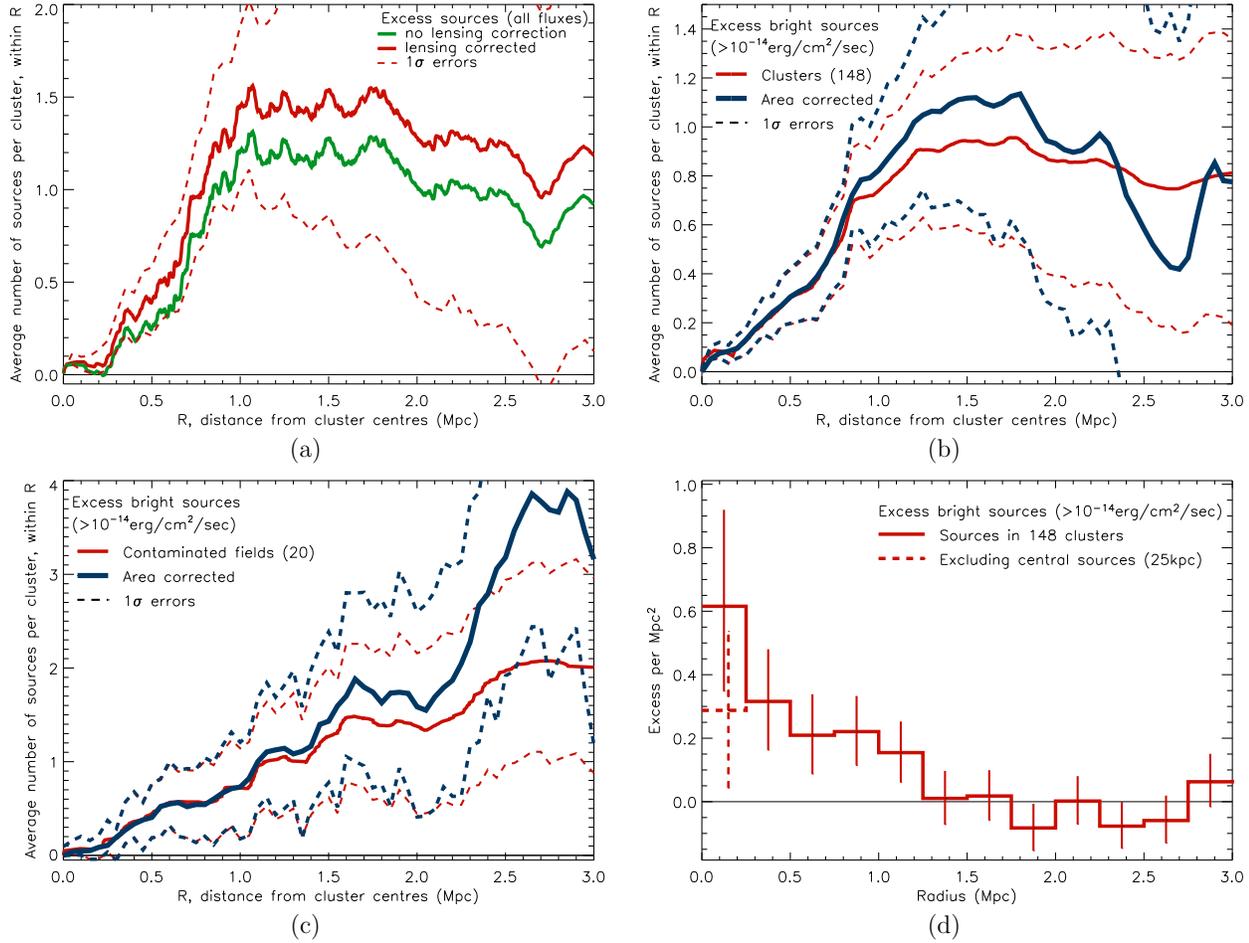}
\caption{ The number of excess sources per field within a given radius
(total detected sources minus the predicted background and foreground
sources), and 1$\sigma$ errors, for {\bf (a)} 148 uncontaminated
cluster fields with $0.1<z<0.9$ corrected for gravitational lensing
(red line and error bars), and without the lensing correction (green
line). All sources are shown, and the cluster fieldss have a
significant excess of sources within 1 Mpc; {\bf (b)} the same 148
clusters (thin red line), but restricted to sources with flux $>
10^{-14}$ erg cm$^{-2}$sec$^{-1}$, which are detectable in almost all
regions of all cluster images. The thick blue `area corrected' line
shows the same distribution corrected for the missing area at each
radius, due to low sensitivity or gaps in the detector.  (The `area
corrected' distribution has higher errors and a more variable line at
high radius, as a lower fraction of the total area is covered in these
regions.); {\bf (c)} 20 contaminated cluster fields i.e. those with
another probable cluster in the field of view. As in plot (b), the
thin red line shows sources $> 10^{-14}$ erg cm$^{-2}$sec$^{-1}$ and
the thick blue `area corrected' line includes a correction for the
missing area at high radius.  The significant excess of sources at $1
- 3$ Mpc, when corrected for the missing area, shows that these fields
contain sources which are not associated with the main cluster, and
therefore should not be included in this analysis. {\bf (d)} The
number density of excess sources in each 0.5 Mpc bin, for the same 148
uncontaminated cluster fields. The dashed line shows the first bin
excluding sources within 25 kpc of the cluster centres. }\label{radl3}
\end{center}
\end{figure*}

The radial distribution of all excess sources in the cluster fields is
shown in Figure \ref{radl3}(a). The cluster fields clearly have an
excess of around 1.5 sources per cluster, whereas the blank fields
have no statistical excess. Although this excess is of low
significance at 3 Mpc, all of the excess sources lie within 1 Mpc from
the cluster centres and within this radius the significance of the
excess is $>3\sigma$, with a maximum significance of $>3.5\sigma$ at
0.85 Mpc. There are no excess sources at $\gtrsim 1$ Mpc despite the
fact that two thirds of all detected sources are beyond this
radius. Figure \ref{radl3}(a) also shows the radial distribution
without correction for gravitational lensing. As explained in Section
\ref{Lensing}, the number of sources per cluster field is lower but
still highly significant.

It is possible that the lack of sources at $>1$ Mpc is due to the fall
off in sensitivity with radius. To check for this the radial
distribution of sources brighter than $10^{-14}$ erg cm$^{-2}$
sec$^{-1}$, which can be well detected at all radii, is plotted in
Figure \ref{radl3}(b). In addition some of the lack of sources could
be due to the reduced sky area at higher radius, so the distribution
for bright sources was corrected for the proportion of missing area at
each increase in radius. Both distributions show that the cluster
sources are still found within $\sim$1 Mpc, with a $\sim 3\sigma$
excess in this area. There is no excess above this radius, although as
the errors are larger in this plot some cluster sources could lie
beyond 1 Mpc. The correction for gravitational lensing for these
brighter sources is not plotted as it negligible (see Section
\ref{Lensing}).

Figure \ref{radl3}(c) shows the same figure for the twenty
contaminated fields, which are those which had a second region of
extended X-ray emission that was not clearly associated with the
cluster, or an optically detected cluster in the field. The
distribution of sources is clearly different to that for the 148
uncontaminated clusters, with excess sources seen up to 3 Mpc from the
cluster centres. The number of sources per cluster in the contaminated
fields is similar at 1 Mpc (within the errors) but larger at higher
radius. This justifies the decision not to include these clusters in
the analysis, as it is very likely that they include sources
associated with the contaminating clusters on the outskirts of the
fields.



Figure \ref{radl3}(d) shows the density of sources with flux $>
10^{-14}$ erg cm$^{-2}$sec$^{-1}$ as a function of radius. It is clear from this
figure that a number of sources lie in the very central regions of the
cluster, and are likely to be AGN in the Brightest Cluster Galaxy
(BCG). Whereas the fraction of radio detected AGN in BCGs is known
\citep{Best07} the number of X-ray detected AGN is not well defined,
and detection is complicated by the extended X-ray
emission. \citeauthor{Ruderman} find that $\sim$10 per cent of their
clusters have a detected X-ray source within 250 kpc of the cD galaxy
in the cluster centre.

In this sample 166 clusters have $0.1<z<1$ and no chip boundaries in
the cluster centre. Twelve of these 166 clusters have X-ray sources
within the central 25 kpc, defined from the centre of the X-ray
emission, whereas only one would be expected randomly. One of these
clusters, 3C 295, has two sources within this radius. Outside this
radius there are very few additional sources compared to the
background prediction. The k-corrected luminosities of these sources
are listed in Table \ref{Cluster_tables}.  To find the proportion of
clusters hosting X-ray detected AGN it is necessary to find the
detection threshold at the centre of each cluster. Neglecting the
second source in 3C 295 the fraction of clusters with AGN with 0.5--8
keV k-corrected luminosity $>L_X$ increases from $2.4^{+1.9}_{-1.2}$
per cent at $L_X = 10^{44}$ erg sec$^{-1}$, to $4.9^{+2.5}_{-1.7}$ per cent at
$10^{43}$ erg sec$^{-1}$ and $7.5^{+4.5}_{-3.7}$ per cent at
$10^{42}$ erg sec$^{-1}$. Below this luminosity most clusters are too bright
to detect AGN and so the statistics are not significant.

When the central sources are excluded in Figure \ref{radl3}(d) the
projected source density is flat or slightly falling until $\sim$1.25
Mpc, where it falls to zero. This is consistent with a random
distribution in projected area, which is naively not the expected
distribution of galaxies in clusters. However, the distribution of
X-ray sources here is consistent with the radial distribution of
cluster galaxies in \cite{Martini_pos}, so it may be that X-ray
sources simply trace the underlying population. This will be
investigated further in the next paper in this series (Gilmour et
al. in prep.).

\subsection{Comparison with previous studies}\label{Comparison_results}
Most of the clusters that have been studied by other authors are also
included in this sample, and in the majority of cases the results of
this analysis agree with the previous results within the errors. The
analyses of many small studies, and the larger studies of
\cite{Martini_pos}, \cite{Branchesi} and \cite{Ruderman}, have been
reproduced as far as possible and are compared below.

Comparing the number of excess sources found by the pipeline in
individual clusters to previous studies, the values are in good
agreement for most clusters; A2104 \citep{Martini}, 3c295
\citep{Cappi}, MS1054-03 \citep{Olivia}, MS0451-03 \citep{Molnar}, six
of the clusters from \cite{Cappelluti} and MRC 1138-262
\citep{Pentericci}. For a few clusters this study produces different
results in the overall number of sources to those found in
\citeauthor{Cappelluti} and \citeauthor{Cappi}, as these authors
present results on a chip-by-chip basis rather than as a radial
analysis. Visual inspection of these fields indicates that their
conclusions are consistent with those from this survey.

Of the eight clusters studied spectroscopically by \cite{Martini_pos},
five are included in this study. The results agree very well (less
than 1$\sigma$ difference) with the \citeauthor{Martini_pos} results,
both in terms of number and radial distribution of the
sources. \citeauthor{Martini_pos} find 17 sources within 1 Mpc in
these five clusters, and 12 within 0.5 Mpc. This study gives 17.5
within 1 Mpc and 9.5 within 0.5 Mpc. At higher radii the errors in this
study become too large to draw any conclusions for 5 fields. In both
studies the AGN are predominantly found in the central regions of
these 5 clusters, with twice as many sources at low radius ($<$ 0.5
Mpc) than at higher radius (0.5 - 1 Mpc). However, from Figure
\ref{radl3}(b) it is clear that this is not the case for the full
sample of 148 clusters -- rather the number of sources at $<$0.5 Mpc
is closer to half the value at higher radius (0.5 - 1 Mpc).  
The central concentration of the \citeauthor{Martini_pos} AGN is
therefore not representative of clusters in general, perhaps because
the \citeauthor{Martini_pos} clusters have particular properties.
This will be investigated further in the next paper in this series
(Gilmour et al., in prep.) when the 148 clusters are split into
subsamples according to redshift and cluster properties.

\cite{Branchesi} performed a statistical analysis of point sources in
18 clusters, of which 15 are in this sample. They find a $1.7\sigma$
excess of bright sources within 1 Mpc ($F_X > 10^{-14}$ erg cm$^{-2}$sec$^{-1}$
). The 15 clusters in this paper have the same
$1.7\sigma$ excess at the bright end of the Log N($>S$) -- Log S
distribution, which also fits with the results for the full sample in
Figure \ref{lognlogs}. 
\citeauthor{Branchesi} find 7(2) cluster(blank) sources at $<$0.5 Mpc,
and 4(3) sources at 0.5--1.0 Mpc and conclude that the majority of
sources are in the central 0.5 Mpc of the cluster.  However, they only
search to the edge of the intra-cluster emission, which gives an
average search radius of 0.8 Mpc.
The results for this study, correcting for missing clusters and
different flux bands, agree with the \citeauthor{Branchesi} values
but the number of sources rises steeply beyond 0.8 Mpc, so the
conclusion that the vast majority of AGN are found within 0.5 Mpc is
not confirmed if larger radii are investigated. As an aside, it is
worth noting that five of the fifteen fields investigated here are
classed as contaminated in this study (morphology type 1c, 2c, or 4 in
Table \ref{Cluster_categories}). In agreement with Section
\ref{Radial_results}, a significant number of sources continue to be
found in these fields up to 3 Mpc from the cluster centres.

\cite{Ruderman} study 51 massive galaxy clusters at $0.3<z<0.7$, and
conclude that the point sources lie predominantly in the central 0.5
Mpc, with a secondary excess at 2 - 3 Mpc. This is significantly
different from the results shown in Figure \ref{radl3}(d). In this
study, using the 25 clusters with published redshift, the excess in
the central 0.5 Mpc is found, but the high significance (8$\sigma$) of
this excess found by \citeauthor{Ruderman} and the secondary excess at
larger radius, are not. One possible explanation of this is that
\citeauthor{Ruderman} measure their excess from the point source
density at $> 4$ Mpc which, as they themselves point out, is lower
than that in the control (blank) fields. If the value implied from
their blank fields is applied to the cluster sample, then the
secondary excess at 2--3 Mpc is no longer significant and the central
excess is of lower significance, in agreement with this study. It
appears that their point source density at large radii is artificially
low due to not subtracting the background sources before scaling to
physical radius. The density of non-cluster sources, when scaled to
the cluster redshift, is dependent on the redshift, and at higher
physical radius only the high redshift fields are used to calculate
the point source density, leading to a lower value. As their point
source list and cluster sample are not yet published it is not
possible to test this further.

\subsection{The brightest sources} 
\begin{figure}
\begin{center}
\includegraphics[width=\columnwidth,clip]{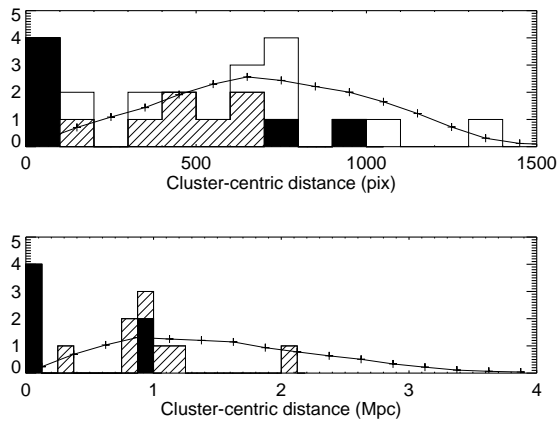}
\caption{The radial distribution of all sources with flux
$>10^{-12.5}$ erg cm$^{-2}$sec$^{-1}$. Solid, hatched and white bars
indicate confirmed cluster AGN, possible cluster AGN and
background/foreground sources respectively. A random distribution,
based on the available sky area, is marked by the solid line. The
confirmed and possible cluster AGN are not randomly distributed in
physical distance from the cluster centres. Ignoring the central four
AGN, a K-S test shows that the distribution in Mpc is not random with
98.5 per cent probability.}\label{highflux}
\end{center}
\end{figure}

As a further test of the validity of the conclusions the distribution
of the very brightest sources, which are clearly highly luminous AGN
if they are in the cluster, was investigated in detail.  Sources
brighter than $10^{-12.5}$ erg cm$^{-2}$sec$^{-1}$, which are easily
detected over the full sample area, were compared to the NED to
identify possible cluster AGN and eliminate contaminating sources. The
results, shown in Figure \ref{highflux}, confirm that the AGN
primarily lie within 1.25 Mpc from the cluster centre. The 13 AGN
which could be cluster members are clustered in two groups - one at
the cluster centres and a second at $\sim$1 Mpc. There is tentative
evidence here that the very brightest sources lie in the outskirts of
the cluster, but this will be investigated in the next paper in this
series.  It is clear that these sources are not foreground objects but
are associated with the clusters, as they are not drawn from a random
distribution with $> 99.93$ per cent probability for all AGN, and 98.5
per cent if the central four AGN are excluded.

\section{Conclusions}
The X-ray point source population in moderately deep ($>10$ ksec)
{\it Chandra} observations of 148 cluster fields and 44 blank fields were
calculated and compared in order to estimate the number of X-ray
sources in the galaxy clusters. The number of sources is found to be
low, with 1.5 sources in a typical cluster. This result is significant
to $>3 \sigma$, but the actual number of sources per cluster depends
on the exposure time and the redshift of the cluster. Over half of
these sources have fluxes corresponding to luminosities
$>10^{42}$ erg sec$^{-1}$ and are likely to be AGN. The population is not
dominated by AGN in the central galaxy (BCG), as only twelve clusters
have a central source, rather the sources are AGN or star-burst
galaxies in normal cluster members. They are all found in the
central 1 Mpc of the cluster, and are randomly distributed in
projected area within this radius.

Many of the clusters covered by similar studies
(e.g. \citealt{Branchesi}; \citealt{Martini_pos}; \citealt{Ruderman}) are
included in this sample, and when the same clusters are compared then
the results are generally in good agreement. However the conclusions
drawn from these papers, which use smaller numbers of clusters, are
often not borne out by this larger study. For example a higher number
of sources is found in the central 0.5 Mpc than the annulus at 0.5 - 1
Mpc in all three previous studies of more than six clusters, and
whilst these results can be reproduced by this study for the smaller
samples, the cluster population in general has more sources in the
outskirts (0.5 - 1 Mpc) of the cluster than the central 0.5 Mpc. In
Appendix A.2 it is demonstrated that samples of less
than five clusters suffer strongly from cosmic variance in the number
of background sources, but larger samples are also affected to some
extent. The question of whether the discrepancies between this study
and some of the previous papers in this field is due to cluster
properties or the larger sample size in this study will be
investigated in the next paper in this series.

This paper describes a sample of point sources in cluster fields which
can be used to investigate the number and properties of X-ray
sources in galaxy clusters as a function of cluster properties and
redshift, and hence increase our understanding of the links between
environment and AGN. This, the first paper in the series, serves as an
introduction to the sample and comparison to previous
studies. Subsequent papers (Gilmour et al. in prep) will use this
study to investigate in more detail the environments of cluster AGN.
\vspace{-1cm}
\section*{Acknowledgments}
\vspace{-0.2cm}
R. Gilmour would like to thank P. Martini for useful discussions and
P. Freeman for providing the source code for {\sc wavdetect}.  P. Best
and O. Almaini would like to thank the Royal Society for generous
financial support through its University Research Fellowship
scheme. This research has made use of the NASA/IPAC Extragalactic
Database (NED) which is operated by the Jet Propulsion Laboratory,
California Institute of Technology, under contract with the National
Aeronautics and Space Administration.
\vspace{-0.5cm}



\begin{onecolumn}

\setlength{\tabcolsep}{2pt}
\scriptsize  
\begin{longtable}{llcccrrrrrrrrc} 
\multicolumn{14}{c}{{\tablename} \thetable{}} \\[0.5ex] \hline \hline\\
NED Name& Obs Ids & RA & DEC & Array & Exp & $F_x$ & $L_x 1$ & $L_x 2$ & r$\chi^2$&z & Ref & Excess & Centre \\ \hline\\
\endfirsthead 
\multicolumn{14}{c}{{\tablename} \thetable{} -- Continued} \\[0.5ex] \hline 
\hline \\NED Name& Obs Ids & RA & DEC & Array & Exp & $F_x$ & $L_x 1$ & $L_x 2$ & r$\chi^2$&z & Ref &Excess & Centre\\ \hline\\
\hline\\ \endhead 
\multicolumn{14}{r}{{Continued on next page\ldots}} \\ \endfoot \\[-5ex]
\\ \endlastfoot \\[-5ex]

\hline\\
\multicolumn{14}{l}{\bf Morphology 1 clusters:} \\ \hline
CL 0016+1609		& 520		& 00:18:33.6 	&  +16:26:12.6	& I	&       67.2	&        2.32 	&       23.06 	&        8.31 	&        0.92	& 0.544	& 1,2	 & $ -2.37^{+ 3.18}_{- 1.94}$& \\
1RXS J003539.8-122247	& 5010		& 00:25:29.7 	&  -12:22:41.9	& I	&       24.7	&        1.03 	&       13.40 	&        5.79 	&        0.95	& 0.584	& 3	 & $ -3.23^{+ 2.32}_{- 0.87}$& \\
ZwCl 0024.0+1652	& 929		& 00:26:35.8 	&  +17:09:41.1	& S	&       38.9	&        0.50 	&        2.71 	&        1.32 	&        0.90	& 0.39	& 4	 & $ 8.92^{+ 5.67}_{- 4.56}$ & \\
RX J0027.6+2616		& 3249		& 00:27:45.4 	&  +26:16:22.5	& I	&       9.9	&        0.83 	&        3.51 	&        1.35 	&        0.96	& 0.367	& 5	 & $ 0.30^{+ 3.18}_{- 1.94}$ & \\
CRSS J0030.5+2618	& 1190,1226	& 00:30:34.0 	&  +26:18:09.9	& S	&       37.1	&        0.26 	&        2.36 	&        0.94 	&        1.02	& 0.50	& 6	 & $ 0.90^{+ 4.12}_{- 2.96}$ & \\
ABELL 0068		& 3250		& 00:37:06.4 	&  +09:09:29.3	& I	&       10.0	&        4.18 	&        7.61 	&        2.87 	&        0.91	& 0.255	& 7	 & $ 4.4^{+ 4.12}_{- 2.96}$  & \\
ABELL 0209		& 522,3579	& 01:31:53.4 	&  -13:36:44.1	& I	&       19.9	&        6.94 	&        8.03 	&        3.58 	&        0.99	& 0.206	& 7	 & $ -5.52^{+ 3.19}_{- 1.95}$& \\
NSCS J015924+003024	& 5777		& 01:59:17.1 	&  +00:30:13.3	& I	&       19.8	&        0.43 	&        2.32 	&        1.30 	&        1.12	& 0.386	& 8	 & $ -2.89^{+ 2.94}_{- 1.66}$& \\
WARP J0216.5-1747	& 5760		& 02:16:32.5 	&  -17:47:34.9	& I	&       36.1	&        0.79 	&       10.30 	&        4.07 	&        1.06	& 0.578	& 9	 & $ 5.47^{+ 4.42}_{- 3.28}$ & \\
CL J023026.6+183622	& 5754		& 02:30:28.6 	&  +18:36:14.7	& I	&       67.6	&        0.14 	&        3.88 	&        1.35 	&        1.02	& 0.799	& 10	 & $ 2.28^{+ 3.96}_{- 2.78}$ & \\
RXC J0232.2-4420	& 4993		& 02:32:18.5 	&  -44:20:48.2	& I	&       18.0	&        6.89 	&       16.55 	&        7.42 	&        1.11	& 0.284	& 11	 & $ -1.23^{+ 3.40}_{- 2.18}$& \\
MACS J0242.6-2132	& 3266		& 02:42:35.9 	&  -21:32:26.3	& I	&       11.7	&        4.72 	&       15.03 	&        7.56 	&        1.00	& 0.314	& 12	 & $ 2.39^{+ 3.78}_{- 2.60}$ & \\
ABELL 0383		& 524,2320	& 02:48:03.5 	&  -03:31:44.4	& I	&       29.2	&        5.46 	&        5.36 	&        2.79 	&        1.19	& 0.187	& 7	 & $ 2.3^{+ 5.34}_{- 4.23}$  & \\
1RXS J025709.6-232549	& 1654		& 02:57:09.1 	&  -23:26:05.2	& I	&       19.8	&        1.77 	&       15.64 	&        6.66 	&        0.95	& 0.505	& 3	 & $ -1.46^{+ 2.93}_{- 1.66}$& \\
CL J030221.3-042329	& 5782		& 03:02:21.1 	&  -04:23:24.5	& I	&       10.0	&        2.21 	&        8.96 	&        4.15 	&        0.83	& 0.35	& 10	 & $ 0.36^{+ 3.18}_{- 1.94}$ & \\
Cl 0302+1658		& 525		& 03:05:31.6 	&  +17:10:08.6	& I	&       10.0	&        0.45 	&        2.84 	&        1.19 	&        0.38	& 0.424	& 2	 & $ -1.12^{+ 2.66}_{- 1.32}$& \\
1RXS J032649.5-004341	& 5810		& 03:26:50.0 	&  -00:43:51.5	& I	&       9.9	&        2.29 	&       15.96 	&        6.48 	&        0.64	& 0.448	& 13	 & $ 0.50^{+ 2.93}_{- 1.66}$ & \\
MACS J0329.7-0212	& 6108		& 03:29:41.6 	&  -02:11:46.6	& I	&       39.4	&        2.12 	&       15.00 	&        6.30 	&        1.05	& 0.45	& 14	 & $ -1.50^{+ 3.40}_{- 2.18}$& \\
CL J033310.2-245641	& 5764		& 03:33:10.5 	&  -24:56:32.5	& I	&       37.0	&        0.15 	&        1.36 	&        0.76 	&        0.91	& 0.475	& 10	 & $ -3.9^{+ 3.18}_{- 1.94}$ & \\
CL J035043.9-380125	& 7227		& 03:50:40.8 	&  -38:02:09.9	& I	&       24.4	&        0.21 	&        0.97 	&        0.55 	&        0.70	& 0.363	& 10	 & $ 0.17^{+ 3.78}_{- 2.60}$ & \\
CL J035559.3-374146	& 5761		& 03:55:59.4 	&  -37:41:45.9	& I	&       27.6	&        0.32 	&        2.73 	&        1.46 	&        0.98	& 0.473	& 10	 & $ -1.91^{+ 3.18}_{- 1.94}$& \\
RBS 0531		& 3270		& 04:17:34.7 	&  -11:54:35.7	& I	&       12.0	&        6.33 	&       39.05 	&       14.67 	&        0.98	& 0.44	& 15	 & $ 2.19^{+ 3.39}_{- 2.18}$ & \\
RX J0439.0+0715		& 526,1449,3583	& 04:39:00.7 	&  +07:16:05.6	& I	&       27.0	&        5.19 	&        7.66 	&        2.86 	&        0.95	& 0.23	& 16	 & $ 1.84^{+ 4.71}_{- 3.58}$ & \\
RX J0439.0+0520		& 527	        & 04:39:02.3 	&  +05:20:44.0	& I	&       9.5	&        3.73 	&        4.55 	&        2.00 	&        0.72	& 0.208	& 16	 & $ 1.81^{+ 3.96}_{- 2.79}$ & \\
MS 0440.5+0204		& 4196		& 04:43:09.9 	&  +02:10:19.7	& 6,7	&       45.1	&        1.55 	&        1.53 	&        0.63 	&        1.22	& 0.19	& 2,17	 & $ 2.7^{+ 6.48}_{- 5.40}$  & \\
RX J0521.1-2530		& 5758		& 05:21:11.7 	&  -25:31:12.5	& I	&       14.9	&        0.01 	&        0.33 	&        0.33 	&        0.09	& 0.581	& 8	 & $ 1.72^{+ 3.39}_{- 2.18}$ & \\
BMW-HRI J052215.8-362453 & 5837		& 05:22:15.4 	&  -36:25:04.5	& I	&       27.6	&        0.17 	&        1.43 	&        0.67 	&        0.70	& 0.472	& 8	 & $ 2.4^{+ 3.96}_{- 2.78}$  & \\
RBS 0653	        & 4994		& 05:28:52.8 	&  -39:28:20.5	& I	&       16.8	&        4.42 	&       10.48 	&        4.66 	&        0.90	& 0.284	& 11	 & $ 4.84^{+ 4.43}_{- 3.28}$ & \\
CL J054250.8-410005	& 914		& 05:42:49.8 	&  -41:00:00.2	& I	&       50.2	&        0.33 	&        4.87 	&        1.84 	&        0.88	& 0.634	& 18	 & $ 2.79^{+ 3.96}_{- 2.78}$ & \\
MACS J0647.7+7015	& 3196,3584	& 06:47:50.3 	&  +70:14:54.6	& I	&       39.0	&        1.68 	&       19.07 	&        6.33 	&        0.83	& 0.584	& 1	 & $ 2.75^{+ 3.96}_{- 2.78}$ & \\
ZwCl 0735.7+7421	& 4197		& 07:41:44.6 	&  +74:14:37.3	& 6,7	&       45.3	&        5.77 	&        7.81 	&        3.92 	&        1.38	& 0.216	& 2	 & $ -7.48^{+ 4.73}_{- 3.60}$& \\
MACS J0744.9+3927	& 3197,3585,6111& 07:44:52.7 	&  +39:27:26.9	& I	&       89.0	&        1.38 	&       24.35 	&        8.57 	&        1.15	& 0.686	& 1	 & $ 3.65^{+ 4.28}_{- 3.12}$ & \\
PKS 0745-19		& 508,2427	& 07:47:31.4 	&  -19:17:41.7	& S	&       36.8	&       57.56 	&       14.87 	&        4.24 	&        1.56	& 0.103	& 19	 & $ -2.7^{+ 5.92}_{- 4.83}$ & \\
ZwCl 0806.5+2822	& 5774		& 08:09:41.9 	&  +28:12:06.9	& I	&       17.7	&        0.79 	&        2.28 	&        1.21 	&        0.93	& 0.30	& 2	 & $ 1.32^{+ 3.96}_{- 2.79}$ & \\
RX J0819.6+6336		& 2199		& 08:19:26.0 	&  +63:37:24.0	& S	&       14.5	&        2.98 	&        1.10 	&        0.60 	&        1.11	& 0.119	& 16	 & $ 0.99^{+ 5.46}_{- 4.36}$ & \\
RX J0820.9+0751		& 1647		& 08:21:02.0 	&  +07:51:48.8	& S	&       8.2	&        1.63 	&        0.51 	&        0.33 	&        1.02	& 0.11	& 20	 & $ 1.86^{+ 5.11}_{- 3.99}$ & \\
ABELL 0665		& 531,3586	& 08:30:58.7 	&  +65:50:31.4	& I	&       38.6	&        9.12 	&       10.99 	&        4.50 	&        1.13	& 0.182	& 7,21	 & $ 4.48^{+ 5.78}_{- 4.68}$ & \\
4C +55.16		& 4940		& 08:34:55.0 	&  +55:34:21.2	& 6,7	&       92.0	&        3.59 	&        6.35 	&        3.29 	&        1.44	& 0.242	& 22	 & $ 0.79^{+ 6.21}_{- 5.12}$ & 43.76\\
2MASX J08425596+2927272	& 2224		& 08:42:55.9 	&  +29:27:25.5	& S	&       29.4	&        3.14 	&        3.39 	&        1.84 	&        1.25	& 0.194	& 2	 & $ -3.56^{+ 5.11}_{- 3.99}$& \\
ABELL 0697	        & 532,4217	& 08:42:57.5 	&  +36:21:56.1	& I	&       23.6	&        7.30 	&       16.61 	&        6.50 	&        1.10	& 0.282	& 7	 & $ 1.23^{+ 4.12}_{- 2.96}$ & \\
RX J0850.1+3604		& 1659		& 08:50:06.7 	&  +36:04:17.1	& I	&       22.1	&        2.94 	&       12.99 	&        5.19 	&        0.93	& 0.378	& 5,16	 & $ 2.68^{+ 4.12}_{- 2.96}$ & \\
ZwCl 0848.5+3341	& 4205		& 08:51:39.0 	&  +33:31:08.0	& S	&       11.4	&        1.03 	&        4.81 	&        2.04 	&        1.15	& 0.380	& 23	 & $ -2.22^{+ 2.94}_{- 1.66}$& \\
MACS J0913.7+4056	& 509		& 09:13:45.4 	&  +40:56:27.6	& S	&       7.8	&        1.46 	&       10.00 	&        4.18 	&        0.87	& 0.442	& 24	 & $ 0.2^{+ 3.18}_{- 1.94}$  & 44.3\\
ABELL 0773		& 533,3588,5006	& 09:17:52.9 	&  +51:43:39.2	& I	&       30.4	&        6.57 	&        8.54 	&        3.87 	&        1.03	& 0.217	& 7,5	 & $ 10.28^{+ 6.8}_{- 4.98}$ & \\
RX J0926.6+1242		& 5838		& 09:26:36.6 	&  +12:43:03.4	& I	&       31.3	&        0.30 	&        2.77 	&        1.41 	&        1.02	& 0.489	& 8	 & $ 3.92^{+ 4.28}_{- 3.12}$ & \\
MACS J0947.2+7623	& 2202		& 09:47:13.0 	&  +76:23:14.2	& I	&       11.7	&        6.30 	&       24.61 	&       11.48 	&        0.91	& 0.35	& 25	 & $ 1.15^{+ 3.40}_{- 2.18}$ & 44.65\\
ZwCl 0947.2+1723	& 3274		& 09:49:51.8 	&  +17:07:08.1	& I	&       14.3	&        3.19 	&       14.28 	&        5.42 	&        0.73	& 0.383	& 5	 & $ 3.98^{+ 3.96}_{- 2.78}$ & \\
ZwCl 0949.6+5207	& 3195		& 09:52:49.3 	&  +51:53:04.9	& S	&       26.5	&        4.35 	&        5.87 	&        3.26 	&        1.01	& 0.214	& 16	 & $ 5.88^{+ 5.88}_{- 4.78}$ & \\
RX J0956.0+4107		& 5759		& 09:56:03.2 	&  +41:07:13.0	& I	&       40.0	&        0.28 	&        3.79 	&        1.81 	&        0.98	& 0.587	& 8	 & $ -4.2^{+ 2.66}_{- 1.33}$ & \\
CL J095819.3+470217	& 5779		& 09:58:19.2 	&  +47:02:03.5	& I	&       25.1	&        0.25 	&        1.40 	&        0.82 	&        1.06	& 0.39	& 26	 & $ 0.89^{+ 3.96}_{- 2.79}$ & \\
ABELL 0907		& 535,3185,3205	& 09:58:21.9 	&  -11:03:50.9	& I	&       104.6	&        8.58 	&        5.31 	&        2.44 	&        1.42	& 0.153	& 27	 & $ 7.60^{+ 7.66}_{- 6.59}$ & \\
MS 1008.1-1224		& 926		& 10:10:32.3 	&  -12:39:34.5	& I	&       43.6	&        2.21 	&        6.00 	&        2.40 	&        1.23	& 0.301	& 2	 & $ 3.93^{+ 4.97}_{- 3.85}$ & \\
ZwCl 1021.0+0426	& 909		& 10:23:39.7 	&  +04:11:09.2	& I	&       45.8	&       10.14 	&       25.66 	&       11.40 	&        1.00	& 0.291	& 28	 & $ 4.78^{+ 5.22}_{- 4.10}$ & \\
ABELL 1068		& 1652		& 10:40:44.6 	&  +39:57:10.2	& S	&       26.7	&       10.63 	&        5.41 	&        3.22 	&        1.50	& 0.138	& 7	 & $ 8.60^{+ 6.74}_{- 5.66}$ & \\
RX J1008.8+0906		& 3252,5009	& 11:08:55.3 	&  +09:05:58.5	& I	&       34.1	&        1.18 	&        8.73 	&        3.77 	&        0.86	& 0.463	& 23	 & $ 0.46^{+ 3.78}_{- 2.60}$ & \\
WARP J1113.0-2615	& 915		& 11:13:05.1 	&  -26:15:38.8	& I	&       104.2	&        0.10 	&        2.46 	&        1.01 	&        1.09	& 0.725	& 9	 & $ 1.9^{+ 4.12}_{- 2.96}$  & \\
ABELL 1204		& 2205		& 11:13:20.4 	&  +17:35:38.7	& I	&       23.5	&        6.78 	&        5.52 	&        3.41 	&        1.24	& 0.171	& 7,5	 & $ 1.89^{+ 5.57}_{- 4.47}$ & \\
RX J1115.8+0129		& 3275		& 11:15:51.9 	&  +01:29:55.6	& I	&       14.6	&        4.84 	&       18.57 	&        7.72 	&        0.93	& 0.38	& 29 (N1)& $ 0.95^{+ 3.40}_{- 2.18}$ & \\
RX J1120.1+4318		& 5771		& 11:20:06.9 	&  +43:18:06.5	& I	&       19.8	&        0.41 	&        6.11 	&        3.00 	&        0.87	& 0.6	& 30	 & $ 1.31^{+ 3.39}_{- 2.18}$ & \\
RX J1130.9+2326		& 1660		& 11:20:57.4 	&  +23:26:33.1	& I	&       70.6	&        0.25 	&        3.30 	&        1.79 	&        0.95	& 0.562	& 8	 & $ 2.5^{+ 4.43}_{- 3.28}$  & \\
MS 1137.5+6624		& 536		& 11:40:22.3 	&  +66:08:16.1	& I	&       117.2	&        0.27 	&        7.21 	&        2.94 	&        0.87	& 0.782	& 31	 & $ -3.48^{+ 3.40}_{- 2.19}$& \\
ABELL 1361		& 2200,3369	& 11:43:39.7 	&  +46:21:20.0	& S	&       15.9	&        4.60 	&        1.64 	&        0.97 	&        1.07	& 0.117	& 7,5	 & $ 0.8^{+ 5.69}_{- 4.59}$  & \\
ABELL 1413		& 537,1661,5003	& 11:55:18.0 	&  +23:24:16.2	& I	&       94.1	&       16.17 	&        8.33 	&        3.65 	&        1.28	& 0.142	& 7,5	 & $ 0.55^{+ 6.59}_{- 5.50}$ & \\
ABELL 1446		& 4975		& 12:02:04.7 	&  +58:02:12.0	& S	&       58.2	&        2.87 	&        0.77 	&        0.44 	&        1.11	& 0.103	& 5	 & $ 0.36^{+ 7.52}_{- 6.46}$ & 41.70\\
CLG J1205+4429		& 4162		& 12:05:51.4 	&  +44:29:10.8	& S	&       29.7	&        0.05 	&        0.82 	&        0.50 	&        1.08	& 0.592	& 32	 & $ -3.88^{+ 2.94}_{- 1.66}$& \\
RXC J1206.2-0848	& 3277		& 12:06:12.4 	&  -08:48:03.9	& I	&       23.4	&        4.70 	&       28.82 	&       10.65 	&        1.07	& 0.441	& 33,29	 & $ 1.41^{+ 3.60}_{- 2.40}$ & \\
RBS 1080		& 5833		& 12:13:23.1 	&  -26:18:07.9	& 6,7	&       9.9	&        0.61 	&        1.41 	&        0.60 	&        0.95	& 0.278	& 34	 & $ -1.75^{+ 3.40}_{- 2.18}$& 43.16\\
RX J1213.5+0253		& 4934		& 12:13:35.0 	&  +02:53:47.9	& I	&       18.7	&        0.25 	&        1.38 	&        0.63 	&        0.77	& 0.409	& 8	 & $ 0.60^{+ 3.60}_{- 2.40}$ & \\
RX J1216.3+2633		& 4931		& 12:16:19.9 	&  +26:33:12.4	& I	&       17.5	&        0.20 	&        1.41 	&        0.80 	&        0.86	& 0.428	& 6	 & $ -1.1^{+ 3.18}_{- 1.94}$ & \\
RX J1221.4+4918		& 1662		& 12:21:26.3 	&  +49:18:27.2	& I	&       78.7	&        0.38 	&        7.47 	&        3.26 	&        0.95	& 0.7	& 6	 & $ 0.41^{+ 3.78}_{- 2.60}$ & \\
CL J122201.9+270919	& 5766		& 12:22:01.9 	&  +27:09:32.4	& I	&       49.0	&        0.18 	&        1.54 	&        0.85 	&        1.08	& 0.472	& 10	 & $ 0.49^{+ 3.96}_{- 2.79}$ & \\
BMW-HRI J122657.3+333253 & 5014		& 12:26:58.0 	&  +33:32:47.1	& I	&       32.6	&        0.86 	&       26.81 	&       10.23 	&        1.04	& 0.89	& 35	 & $ 1.23^{+ 3.39}_{- 2.18}$ & \\
ABELL 3541	        & 1648		& 13:03:42.4 	&  -24:14:45.3	& S	&       9.7	&        7.78 	&        3.35 	&        1.80 	&        1.28	& 0.128	& 7	 & $ -4.45^{+ 3.80}_{- 2.62}$& \\
MACS J1311.0-0311	& 3258,6110	& 13:11:01.7 	&  -03:10:38.5	& I	&       77.8	&        1.11 	&        9.92 	&        4.79 	&        0.95	& 0.49	& 13	 & $ 4.61^{+ 4.71}_{- 3.57}$ & \\
ABELL 1689		& 7289		& 13:11:29.5 	&  -01:20:29.8	& I	&       74.9	&       18.52 	&       16.40 	&        7.10 	&        1.42	& 0.183	& 7	 & $ 5.92^{+ 6.57}_{- 5.48}$ & \\
RX J1320.0+7003		& 3278		& 13:20:07.9 	&  +70:04:36.8	& I	&       20.5	&        1.74 	&        5.74 	&        2.58 	&        0.73	& 0.328	& 7,5	 & $ 5.42^{+ 4.57}_{- 3.43}$ & \\
ZwCl 1332.8+5043	& 5772		& 13:34:20.1 	&  +50:31:01.2	& I	&       17.6	&        0.36 	&        5.54 	&        2.67 	&        1.00	& 0.62	& 30	 & $ 3.71^{+ 3.78}_{- 2.59}$ & \\
RX J1340.5+4017		& 3223		& 13:40:32.9 	&  +40:17:38.7	& S	&       46.4	&        0.16 	&        0.13 	&        0.13 	&        1.05	& 0.171	& 36	 & $ 6.78^{+ 7.18}_{- 6.10}$ & \\
LCDCS 0829		& 3592		& 13:47:30.8 	&  -11:45:10.1	& I	&       57.4	&       10.78 	&       68.09 	&       23.48 	&        1.16	& 0.451	& 37,38	 & $ -1.92^{+ 3.40}_{- 2.18}$& \\
RDCS J1350+6007		& 2229		& 13:50:48.3 	&  +60:07:06.0	& I	&       58.1	&        0.12 	&        3.99 	&        1.98 	&        1.05	& 0.804	& 39	 & $ 2.70^{+ 4.12}_{- 2.96}$ & \\
ZwCl 1358.1+6245	& 516		& 13:59:50.6 	&  +62:31:02.9	& S	&       52.1	&        2.42 	&        8.31 	&        3.80 	&        1.11	& 0.328	& 2	 & $ 1.52^{+ 5.34}_{- 4.23}$ & \\
ABELL 1835		& 495,496	& 14:01:02.0 	&  +02:52:41.6	& S	&       30.2	&       18.40 	&       34.36 	&       14.89 	&        1.58	& 0.253	& 7	 & $ -2.33^{+ 4.58}_{- 3.44}$& \\
3C 295			& 578		& 14:11:20.4 	&  +52:12:10.0	& S	&       17.9	&        0.87 	&        6.92 	&        3.43 	&        1.21	& 0.46	& 40	 & $ 7.57^{+ 4.84}_{- 3.71}$ & 43.77(N2)\\
NSCS J141623+444558	& 541		& 14:16:27.9 	&  +44:46:44.5	& I	&       30.7	&        0.58 	&        3.38 	&        1.80 	&        1.02	& 0.4	& 6	 & $ -2.53^{+ 3.40}_{- 2.18}$& \\
MACS J1423.8+2404	& 4195		& 14:23:47.9 	&  +24:04:42.6	& 6,7	&       115.2	&        1.92 	&       22.14 	&        9.91 	&        1.20	& 0.545	& 1	 & $ 3.73^{+ 5.10}_{- 3.98}$ & \\
ABELL 1914		& 542,3593	& 14:26:02.0 	&  +37:49:32.8	& I	&       26.7	&       19.03 	&       14.54 	&        6.18 	&        1.00	& 0.171	& 7	 & $ 5.23^{+ 5.78}_{- 4.68}$ & \\
RBS 1460		& 5793		& 15:04:07.5 	&  -02:48:16.1	& I	&       38.9	&       25.15 	&       32.46 	&       13.57 	&        1.22	& 0.215	& 41	 & $ -1.19^{+ 4.72}_{- 3.58}$& \\
ABELL 2034		& 2204		& 15:10:11.8 	&  +33:30:54.3	& I	&       53.8	&       10.73 	&        3.41 	&        1.59 	&        0.97	& 0.113	& 7	 & $ 3.82^{+ 8.49}_{- 7.43}$ & \\
RX J1532.5+3021		& 1665		& 15:32:53.8 	&  +30:20:58.6	& I	&       9.9	&        5.92 	&       22.92 	&       11.18 	&        1.08	& 0.345	& 16 (N3)& $ 1.18^{+ 3.40}_{- 2.18}$ & \\
ABELL 2111		& 544		& 15:39:41.3 	&  +34:25:06.7	& I	&       10.3	&        3.57 	&        5.18 	&        2.26 	&        1.02	& 0.229	& 7	 & $ -1.70^{+ 3.40}_{- 2.19}$& \\
ABELL 2104		& 895		& 15:40:07.9 	&  -03:18:17.5	& S	&       49.0	&        8.94 	&        5.42 	&        1.80 	&        1.72	& 0.153	& 7,42	 & $ 12.18^{+ 7.41}_{- 6.34}$& \\
WARP J1552.2+2013	& 3214		& 15:52:12.8 	&  +20:13:39.9	& S	&       14.9	&        0.23 	&        0.11 	&        0.08 	&        1.06	& 0.136	& 6	 & $ -3.30^{+ 4.86}_{- 3.74}$& \\
MACS J1621.4+3810	& 3594,6109,6172& 16:21:24.7 	&  +38:10:08.5	& I	&       77.2	&        1.50 	&       11.50 	&        5.45 	&        1.03	& 0.465	& 43	 & $ 0.12^{+ 3.96}_{- 2.79}$ & \\
ABELL 2204		& 499	        & 16:32:46.9 	&  +05:34:31.5	& S	&       10.0	&       32.58 	&       19.81 	&        8.68 	&        1.24	& 0.152	& 7	 & $ 1.89^{+ 4.58}_{- 3.44}$ & \\
ABELL 2218		& 553,1454,1666	& 16:35:51.5 	&  +66:12:36.8	& I	&       60.3	&        5.76 	&        4.77 	&        2.11 	&        0.96	& 0.176	& 7	 & $ 7.1^{+ 6.92}_{- 5.84}$  & \\
ABELL 2219		& 896		& 16:40:19.9 	&  +46:42:35.3	& S	&       42.1	&       16.04 	&       22.61 	&        9.30 	&        1.57	& 0.226	& 7	 & $ -7.51^{+ 4.30}_{- 3.15}$& \\
Hercules A		& 6257		& 16:51:08.2 	&  +04:59:33.0	& 6,7	&       49.3	&        4.34 	&        2.77 	&        1.40 	&        1.16	& 0.154	& 44	 & $ 2.96^{+ 6.75}_{- 5.67}$ & \\
ABELL 2256		& 3245		& 17:20:08.4 	&  +27:40:11.0	& I	&       10.0	&        6.37 	&        4.58 	&        2.14 	&        0.99	& 0.164	& 7	 & $ 2.35^{+ 4.44}_{- 3.29}$ & \\
SDSS-C4 3072		& 1453,3224,4361& 17:20:10.1 	&  +26:37:30.8	& I	&       54.4	&       14.37 	&       10.30 	&        4.69 	&        1.22	& 0.164	& 16	 & $ 1.2^{+ 6.20}_{- 5.11}$  & \\
MACS J1720.2+3536	& 6107	        & 17:20:16.7 	&  +35:36:23.9	& I	&       33.5	&        2.59 	&       13.05 	&        5.69 	&        1.07	& 0.391	& 43	 & $ 0.93^{+ 3.96}_{- 2.78}$ & \\
ABELL 2294		& 3246		& 17:24:12.0 	&  +85:53:10.8	& I	&       9.6	&        7.97 	&        6.48 	&        2.23 	&        1.06	& 0.178	& 7	 & $ 9.13^{+ 5.22}_{- 4.10}$ & \\
MS 2053.7-0449		& 551,1667	& 20:56:21.1 	&  -04:37:46.3	& I	&       88.4	&        0.21 	&        2.72 	&        1.13 	&        0.96	& 0.583	& 2	 & $ 1.83^{+ 4.43}_{- 3.28}$ & \\
MACS J2129.4-0741	& 3199,3595	& 21:29:26.0 	&  -07:41:28.2	& I	&       36.7	&        1.61 	&       18.25 	&        6.57 	&        1.00	& 0.57	& 1	 & $ 6.77^{+ 4.57}_{- 3.43}$ & \\
RBS 1748		& 552		& 21:29:40.0 	&  +00:05:19.7	& I	&       9.9	&        6.72 	&        9.54 	&        4.24 	&        0.98	& 0.224	& 16	 & $ 0.70^{+ 3.79}_{- 2.60}$ & \\
MS 2137.3-2353		& 928		& 21:40:15.2 	&  -23:39:40.1	& S	&       39.4	&        3.69 	&       11.63 	&        5.54 	&        1.34	& 0.313	& 2	 & $ -4.44^{+ 4.44}_{- 3.30}$& \\
ABELL 2390		& 4193		& 21:53:39.0 	&  +17:41:15.3	& 6,7	&       93.7	&       16.19 	&       23.69 	&        9.01 	&        1.66	& 0.23	& 7,45	 & $ 12.22^{+ 7.0}_{- 5.92}$ & 42.90\\
ABELL 2409		& 3247		& 22:00:52.9 	&  +20:58:22.3	& I	&       10.2	&        8.75 	&        5.00 	&        2.15 	&        0.97	& 0.148	& 7	 & $ -3.44^{+ 3.41}_{- 2.20}$& \\
1RXS J221144.6-034947	& 3284		& 22:11:45.9 	&  -03:49:46.6	& I	&       17.7	&        7.06 	&       14.29 	&        5.13 	&        1.04	& 0.27	& 41	 & $ 7.5^{+ 4.84}_{- 3.71}$  & \\
MACS J2214.9-1359	& 3259		& 22:14:57.3 	&  -14:00:12.3	& I	&       19.2	&        1.92 	&       15.07 	&        5.98 	&        1.09	& 0.483	& 46	 & $ 2.81^{+ 3.78}_{- 2.60}$ & \\
MACS J2229.8-2756	& 3286		& 22:29:45.2 	&  -27:55:36.4	& I	&       16.0	&        3.22 	&       11.12 	&        6.36 	&        0.90	& 0.322	& 47	 & $ 2.13^{+ 3.96}_{- 2.78}$ & \\
1RXS J224322.6-093549	& 3260		& 22:43:20.9 	&  -09:35:42.8	& I	&       20.3	&        2.72 	&       17.42 	&        7.02 	&        1.06	& 0.439	& 48	 & $ 1.52^{+ 3.60}_{- 2.40}$ & \\
1RXS J224505.2+263758	& 3287		& 22:45:04.7 	&  +26:38:03.5	& I	&       16.1	&        3.36 	&        9.52 	&        4.02 	&        1.05	& 0.304	& 5	 & $ 0.83^{+ 3.40}_{- 2.18}$ & \\
RX J2247.2+0337		& 911		& 22:47:28.0 	&  +03:37:00.8	& I	&       48.7	&        0.12 	&        0.11 	&        0.08 	&        1.02	& 0.18	& 6	 & $ 14.40^{+ 7.48}_{- 6.41}$& \\
ABELL S1063		& 4966		& 22:48:44.8 	&  -44:31:46.4	& I	&       26.6	&       11.12 	&       40.28 	&       16.32 	&        1.09	& 0.348	& 29	 & $ 7.30^{+ 4.71}_{- 3.57}$ & \\
RBS 1906		& 5769		& 22:51:47.5 	&  -32:06:12.5	& 6,7	&       10.3	&        0.32 	&        0.53 	&        0.21 	&        0.74	& 0.246	& 41	 & $ 7.40^{+ 5.9}_{- 3.97}$  &44.00 \\
ABELL S107		& 1562		& 22:58:48.2 	&  -34:48:07.3	& S	&       72.1	&        4.74 	&       13.72 	&        5.69 	&        2.18	& 0.31	& 49,50	 & $ 2.70^{+ 5.68}_{- 4.57}$ & \\
ABELL 2537		& 4962		& 23:08:22.1 	&  -02:11:27.9	& 6,7	&       36.0	&        2.85 	&        7.27 	&        2.88 	&        1.19	& 0.295	& 51	 & $ -3.29^{+ 4.44}_{- 3.29}$& 42.04\\
ABELL 2631		& 3248		& 23:37:38.7 	&  +00:16:08.5	& I	&       9.1	&        4.12 	&        9.19 	&        3.78 	&        1.05	& 0.278	& 7,29	 & $ 4.65^{+ 4.12}_{- 2.96}$ & \\
ABELL 2667		& 2214		& 23:51:39.3 	&  -26:05:03.5	& S	&       9.6	&       11.90 	&       17.72 	&        8.71 	&        1.14	& 0.226	& 52,29	 & $ 0.89^{+ 3.79}_{- 2.61}$ & \\
\hline\\
\multicolumn{14}{l}{\bf Morphology 2 clusters:} \\ \hline
RX J0404.6+1109		& 3269		& 04:04:32.9 	&  +11:08:08.0	& I	&       21.7	&        0.89 	&        3.42 	&        1.14 	&        0.95	& 0.355	& 5	 & $ 5.69^{+ 4.43}_{- 3.28}$ & \\
RX J0853.2+5759		& 5765		& 08:53:16.8 	&  +57:59:44.4	& I	&       24.5	&        0.16 	&        1.40 	&        0.72 	&        0.80	& 0.475	& 6	 & $ 2.73^{+ 3.96}_{- 2.78}$ & \\
RX J1006.9+3200		& 5819		& 10:06:54.5 	&  +32:01:32.6	& I	&       10.8	&        1.45 	&        7.58 	&        3.51 	&        1.08	& 0.398	& 54	 & $ 1.68^{+ 3.39}_{- 2.18}$ & \\
ZwCl 1006.1+1201	& 925		& 10:08:47.5 	&  +11:47:36.1	& I	&       29.3	&        2.63 	&        3.61 	&        1.61 	&        1.25	& 0.221	& 2	 & $ 7.61^{+ 5.56}_{- 4.45}$ & \\
ABELL 1201		& 4216		& 11:12:54.6 	&  +13:26:02.4	& S	&       34.2	&        3.21 	&        2.51 	&        1.29 	&        1.22	& 0.169	& 7	 & $ -7.50^{+ 5.24}_{- 4.13}$& \\
ZwCl 1112.2+5318	& 5008		& 11:15:15.8 	&  +53:19:54.3	& I	&       18.0	&        1.56 	&       11.65 	&        5.50 	&        0.87	& 0.466	& 23	 & $ 1.76^{+ 3.60}_{- 2.40}$ & \\
ABELL 1240		& 4961		& 11:23:37.3 	&  +43:06:08.3	& I	&       51.1	&        0.93 	&        1.01 	&        0.54 	&        0.97	& 0.196	& 55	 & $ -3.51^{+ 5.80}_{- 4.70}$& \\
NSCS J125606+255746	& 3212		& 12:56:02.5 	&  +25:56:38.0	& S	&       26.9	&        0.13 	&        0.22 	&        0.14 	&        0.91	& 0.232	& 6,56	 & $ -5.31^{+ 4.59}_{- 3.45}$& \\
ABELL 1763		& 3591		& 13:35:18.3 	&  +41:00:00.7	& I	&       19.6	&        5.85 	&        8.08 	&        3.78 	&        1.26	& 0.223	& 57	 & $ -1.78^{+ 3.96}_{- 2.79}$& \\
RX J1354.2-0221		& 5835		& 13:54:17.1 	&  -02:21:52.6	& I	&       37.6	&        0.14 	&        1.85 	&        1.01 	&        0.87	& 0.546	& 8	 & $ -1.13^{+ 3.40}_{- 2.18}$& \\
MS 1621.5+2640		& 546		& 16:23:35.2 	&  +26:34:21.3	& I	&       29.9	&        1.10 	&        6.54 	&        2.60 	&        1.02	& 0.426	& 2	 & $ 0.94^{+ 3.96}_{- 2.78}$ & \\
RX J1716.4+6708		& 548		& 17:16:48.9 	&  +67:08:25.4	& I	&       51.6	&        0.32 	&        8.82 	&        3.18 	&        0.76	& 0.813	& 58	 & $ 4.49^{+ 4.28}_{- 3.12}$ & \\
ABELL 2261		& 550,5007	& 17:22:27.2 	&  +32:07:57.0	& I	&       33.3	&        9.02 	&       12.53 	&        5.16 	&        1.08	& 0.224	& 7	 & $ -6.87^{+ 3.61}_{- 2.42}$& \\
MACS J1824.2+4309	& 3255		& 18:24:18.5 	&  +43:09:54.2	& I	&       14.9	&        0.03 	&        0.41 	&        0.36 	&        0.07	& 0.487	& 43	 & $ 0.39^{+ 3.18}_{- 1.93}$ & \\
MACS J2228.5+2036	& 3285		& 22:28:33.2 	&  +20:37:12.9	& I	&       19.8	&        2.99 	&       16.25 	&        6.30 	&        0.81	& 0.412	& 5	 & $ 2.52^{+ 3.78}_{- 2.60}$ & \\
ABELL 2550		& 2225		& 23:11:35.7 	&  -21:44:46.8	& S	&       58.6	&        1.03 	&        0.42 	&        0.34 	&        1.23	& 0.123	& 59	 & $ -4.40^{+ 6.78}_{- 5.70}$& \\
\hline\\
\multicolumn{14}{l}{\bf Morphology 3 clusters:} \\ \hline \vspace{0.1cm} 
ABELL 2744		& 2212		& 00:14:19.1 	&  -30:23:23.1	& 6,7	&       24.7	&        8.18 	&       23.31 	&        9.55 	&        1.25	& 0.308	& 7	 & $ 0.99^{+ 4.28}_{- 3.13}$& \\
ABELL 0115		& 3233		& 00:55:50.5 	&  +26:24:35.5	& I	&       49.6	&        2.98 	&        3.28 	&        1.65 	&        1.02	& 0.197	& 7	 & $ 4.32^{+ 6.9}_{- 5.0}$	 & \\
EDCC 586		& 5778		& 01:41:32.6 	&  -30:34:42.6	& I	&       29.3	&        0.17 	&        0.14 	&        0.13 	&        1.23	& 0.168	& 60	 & $ 5.96^{+ 6.29}_{- 5.20}$& \\
CL J0152.7-1357		& 913		& 01:52:44.4 	&  -13:57:17.4	& I	&       36.3	&        0.19 	&        5.12 	&        1.94 	&        0.99	& 0.831	& 61	 & $ 1.31^{+ 3.39}_{- 2.18}$& \\
ABELL 0521		& 901		& 04:54:06.6 	&  -10:13:09.4	& I	&       38.5	&        3.50 	&        6.48 	&        2.72 	&        1.09	& 0.253	& 62	 & $ 2.62^{+ 4.97}_{- 3.85}$& \\
ABELL 0520		& 528,4215	& 04:54:10.1 	&  +02:54:40.7	& I	&       75.5	&        6.38 	&        7.14 	&        2.81 	&        1.20	& 0.203	& 7,5	 & $ 3.25^{+ 6.19}_{- 5.10}$& \\
1RXS J065830.3-555702	& 554,3184	& 06:58:30.3 	&  -55:56:34.6	& I	&       112.4	&       13.22 	&       32.58 	&       11.24 	&        1.14	& 0.296	& 63	 & $ 7.70^{+ 5.67}_{- 4.56}$& \\
CXOU J091554+293316	& 4209		& 09:15:52.5 	&  +29:33:23.6	& I	&       19.1	&        0.08 	&        0.83 	&        0.42 	&        0.85	& 0.500	& 64	 & $ 1.8^{+ 3.39}_{- 2.18}$ & 45.09\\
RXC J1234.2+0947	& 539		& 12:34:21.7 	&  +09:46:56.9	& I	&       9.1	&        2.43 	&        3.66 	&        1.83 	&        0.79	& 0.229	& 5	 & $ 0.77^{+ 3.40}_{- 2.19}$& \\
ABELL 1682		& 3244		& 13:06:50.5 	&  +46:33:25.3	& I	&       8.9	&        4.51 	&        6.83 	&        2.86 	&        1.12	& 0.234	& 7	 & $ 0.82^{+ 3.60}_{- 2.40}$& \\
ABELL 1758		& 2213		& 13:32:42.8 	&  +50:32:55.0	& 6,7	&       55.8	&        7.49 	&       17.28 	&        7.26 	&        1.43	& 0.28	& 7,5	 & $ 0.70^{+ 5.23}_{- 4.11}$& \\
ABELL 2069		& 4965		& 15:24:08.7 	&  +29:53:00.3	& I	&       52.2	&        3.19 	&        1.09 	&        0.56 	&        1.08	& 0.116	& 7,65	 & $ -4.82^{+ 7.13}_{- 6.6}$& \\
FIRST J234229.5+001845	& 5786		& 23:43:41.9 	&  +00:18:07.5	& I	&       29.7	&        1.49 	&        3.22 	&        1.42 	&        1.08	& 0.27	& 13	 & $ 2.63^{+ 4.84}_{- 3.71}$& \\
\hline \vspace{0.2cm} \\ 
\multicolumn{14}{l}{\bf Morphology 1c clusters:} \\ \hline
ABELL 0267		& 1448,3580	& 01:52:42.2 	&  +01:00:40.5	& I	&       27.4	&        4.16 	&        6.27 	&        2.81 	&        1.13	& 0.23	& 7,5	 & $ 3.14^{+ 4.97}_{- 3.85}$ & \\
MACS J0159.8-0849	& 3265,6106	& 01:59:49.3 	&  -08:49:59.9	& I	&       52.8	&        4.18 	&       21.81 	&        9.66 	&        1.06	& 0.4	& 48	 & $ 2.16^{+ 4.12}_{- 2.96}$ & \\
RDCS 0337.4-3457	& 6264		& 03:37:25.0 	&  -34:57:18.6	& I	&       12.2	&        0.02 	&        1.30 	&        1.06 	&        0.26	& 0.84	& 66	 & $ 0.10^{+ 2.65}_{- 1.32}$ & \\
CL J034051.6-282310	& 5780		& 03:40:52.9 	&  -28:23:08.3	& I	&       24.6	&        0.37 	&        1.53 	&        0.94 	&        1.02	& 0.346	& 10	 & $ 2.2^{+ 4.28}_{- 3.12}$  & \\
RX J0451.9+0006		& 5815		& 04:51:54.4 	&  +00:06:19.2	& I	&       10.2	&        1.36 	&        8.27 	&        3.05 	&        0.81	& 0.43	& 67	 & $ 3.34^{+ 3.59}_{- 2.39}$ & \\
MACS J0454.1-0300	& 902		& 04:54:11.2 	&  -03:00:51.3	& S	&       43.5	&        2.37 	&       25.07 	&        9.06 	&        1.08	& 0.55	& 1,31	 & $ -4.82^{+ 2.94}_{- 1.66}$& \\
MACS J0717+3745		& 1655,4200	& 07:17:31.3 	&  +37:45:29.5	& I	&       78.7	&        3.69 	&       35.91 	&       11.53 	&        1.17	& 0.548	& 1	 & $ 3.80^{+ 4.28}_{- 3.12}$ & \\
ABELL 0586		& 530		& 07:32:20.3 	&  +31:37:56.2	& I	&       10.0	&        7.14 	&        5.51 	&        2.27 	&        1.02	& 0.171	& 7	 & $ 0.96^{+ 4.13}_{- 2.97}$ & \\
ABELL 0611		& 3194		& 08:00:56.7 	&  +36:03:23.2	& S	&       35.9	&        3.82 	&        9.62 	&        4.18 	&        1.22	& 0.288	& 7	 & $ 2.99^{+ 5.22}_{- 4.10}$ & 43.29\\
MS 0906.5+1110		& 924		& 09:09:12.8 	&  +10:58:33.0	& I	&       29.6	&        4.69 	&        3.85 	&        1.73 	&        0.94	& 0.175	& 5,2,7	 & $ 1.40^{+ 5.22}_{- 4.11}$ & \\
ABELL 0963		& 903		& 10:17:03.6 	&  +39:02:53.4	& S	&       36.2	&        7.98 	&        9.54 	&        4.53 	&        1.16	& 0.206	& 7,5	 & $ 7.75^{+ 6.38}_{- 5.29}$ & \\
MS 1054-03		& 512		& 10:56:59.0 	&  -03:37:35.0	& S	&       85.7	&        0.63 	&       18.13 	&        6.77 	&        1.04	& 0.823	& 31,68	 & $ 0.42^{+ 4.13}_{- 2.96}$ & \\
MACS J1149.5+2223	& 1656,3589	& 11:49:35.0 	&  +22:24:06.7	& I	&       38.5	&        2.20 	&        1.82 	&        0.83 	&        1.06	& 0.176	& 5,1	 & $ 4.61^{+ 6.19}_{- 5.10}$ & \\
CL J131219.4+390058	& 5781		& 13:12:19.5 	&  +39:00:53.4	& I	&       25.1	&        0.23 	&        1.46 	&        0.93 	&        1.30	& 0.404	& 10	 & $ -3.52^{+ 2.94}_{- 1.66}$& \\
RDCS J1317+2911		& 2228		& 13:17:21.5 	&  +29:11:16.2	& I	&       111.0	&        0.01 	&        0.58 	&        0.31 	&        0.72	& 0.805	& 39	 & $ 4.2^{+ 4.71}_{- 3.58}$  & \\
MS 1455.0+2232		& 543,4192	& 14:57:15.0 	&  +22:20:34.7	& I	&       101.4	&        5.97 	&       11.96 	&        6.04 	&        1.28	& 0.258	& 28	 & $ -5.33^{+ 4.73}_{- 3.60}$& \\
ABELL 2163		& 545,1653	& 16:15:45.8 	&  -06:09:01.8	& I	&       80.3	&       24.45 	&       25.75 	&        7.58 	&        1.65	& 0.203	& 69,70	 & $ -6.78^{+ 4.30}_{- 3.15}$& \\
1RXS J201127.9-572507	& 4995		& 20:11:27.1 	&  -57:25:10.1	& I	&       23.9	&        1.55 	&        3.75 	&        1.91 	&        0.77	& 0.279	& 41	 & $ 1.59^{+ 4.28}_{- 3.13}$ & \\
\hline\\
\multicolumn{14}{l}{\bf Morphology 2c clusters:} \\ \hline
ABELL 1300		& 3276		& 11:31:54.8 	&  -19:55:49.5	& I	&       13.8	&        4.47 	&       12.36 	&        4.74 	&        0.90	& 0.307	& 7,71	 & $ 3.11^{+ 3.96}_{- 2.78}$ & \\
ABELL 1942		& 3290		& 14:38:21.9 	&  +03:40:09.7	& I	&       56.7	&        1.14 	&        1.56 	&        0.73 	&        0.96	& 0.22	& 72	 & $ 7.4^{+ 6.18}_{- 5.9}$   & 41.93\\
\hline\\
\multicolumn{14}{l}{\bf High redshift (proto)clusters:} \\ \hline
XLSSC 029		& 7185		& 02:24:04.1 	&  -04:13:30.1	& 6,7	&       32.9	&   --- 	&       --- 	&       --- 	&       ---	& 1.05	& 73	 & $ 2.20^{+ 3.78}_{- 2.60}$& 43.24\\
RCS J0439-2904		& 3577		& 04:39:38.1 	&  -29:04:55.8	& S	&       86.6	&   --- 	&       --- 	&       --- 	&       ---	& 0.951	& 53	 & $ 0.19^{+ 3.60}_{- 2.40}$& \\
CL J0442+0202		& 3242		& 04:42:23.8 	&  +02:02:19.6	& I	&       43.4	&   --- 	&       --- 	&       --- 	&       ---	& 1.11	& 74	 & $ 6.15^{+ 4.27}_{- 3.12}$& 44.75\\
3C 184			& 3226		& 07:39:28.4 	&  +70:23:39.8	& S	&       18.8	&   --- 	&       --- 	&       --- 	&       ---	& 0.996	& 75	 & $ 0.67^{+ 3.18}_{- 1.93}$& \\
3C 210			& 5821		& 08:58:10.0 	&  +27:50:53.5	& 6,7	&       20.5	&   --- 	&       --- 	&       --- 	&       ---	& 1.169	& 76	 & $ 0.50^{+ 2.93}_{- 1.66}$& 44.06\\
RDCS J0910+5422		& 2227,2452	& 09:10:44.6 	&  +54:22:03.8	& I	&       168.8	&   --- 	&       --- 	&       --- 	&       ---	& 1.11	& 77	 & $ 9.24^{+ 5.22}_{- 4.10}$& \\
CL J100207.7+685848	& 5773	        & 10:02:09.2 	&  +68:58:38.0	& I	&       19.8	&   --- 	&       --- 	&       --- 	&       ---	& 0.928	& 10	 & $ 4.41^{+ 3.78}_{- 2.59}$& \\
PKS 1138-26		& 898		& 11:40:48.3 	&  -26:29:09.9	& S	&       32.5	&   --- 	&       --- 	&       --- 	&       ---	& 2.16	& 78	 & $ 4.83^{+ 4.28}_{- 3.12}$& 45.58\\
RDCS J1252-2927		& 4198,4403	& 12:52:54.5 	&  -29:27:17.1	& I	&       188.2	&   --- 	&       --- 	&       --- 	&       ---	& 1.23	& 79	 & $ 4.92^{+ 4.71}_{- 3.58}$& \\
3C 280			& 2210		& 12:56:58.2 	&  +47:20:22.2	& S	&       45.7	&   --- 	&       --- 	&       --- 	&       ---	& 0.996	& 44,8	 & $ 2.25^{+ 4.12}_{- 2.96}$& \\
3C 294			& 3207,3445	& 14:06:44.0 	&  +34:11:26.1	& 6,7	&       190.9	&   --- 	&       --- 	&       --- 	&       ---	& 1.78	& 76	 & $ 2.90^{+ 4.73}_{- 3.60}$& 44.48\\
WARP J1415.1+3612	& 4163		& 14:15:11.0 	&  +36:12:03.5	& I	&       88.8	&   --- 	&       --- 	&       --- 	&       ---	& 1.03	& 80	 & $ 2.82^{+ 4.12}_{- 2.96}$& \\
3C 324			& 326		& 15:49:48.8 	&  +21:25:37.5	& S	&       39.4	&   --- 	&       --- 	&       --- 	&       ---	& 1.21	& 44	 & $ 4.83^{+ 4.28}_{- 3.12}$& 43.82\\
4C +15.55		& 3229		& 16:25:14.4 	&  +15:45:22.8	& I	&       51.0	&   --- 	&       --- 	&       --- 	&       ---	& 1.406	& 81     & $ 3.95^{+ 3.95}_{- 2.78}$& 45.67\\
\hline \\

\caption{The final cluster sample, split by morphological class.}\label{Cluster_tables}
\end{longtable}
\noindent Notes for Table \ref{Cluster_tables}. N1 - the NED gives z=0.35, but
the X-ray spectral data has a far better fit with z=0.38. N2 - this
cluster has two X-ray point sources within 25kpc, which overlap
slightly. The second, at ~10kpc, has log luminosity $\sim$43.72. N3 -
the NED gives two redshifts, but only one is given in the paper which
NED refers to.  Columns are: {\it NED Name} - Name of cluster in the
NED. If more then one cluster name exists then the nearest is
given. If there is no cluster within 2$\arcmin$ then the nearest
object name at the cluster redshift is given; {\it Obs Ids} - Chandra
Observation ID; {\it RA \& DEC} - Position of cluster as determined
from the X-ray emission (J2000); {\it Array} - ACIS Detector or CCDs
used; {\it Exp} - Good exposure time in ksec, after filtering (average
over the selected chips); {\it $F_x$} - Cluster approximate observed
frame 0.5-8keV flux ($ 10^{-12}$erg/cm$^2$/sec). All fluxes and
luminosities are approximate as the detector response was only
calculated for the central pixel; {\it $L_x 1$} - Cluster approximate
rest frame 0.5-8keV luminosity ($ 10^{44}$erg/sec); {\it $L_x 2$} -
Cluster approximate rest frame 0.1-2.4keV luminosity ($
10^{44}$erg/sec); {\it r$\chi^2$} - reduced $\chi^2$ of the XSPEC fit
to the cluster spectrum; {\it z} - Cluster redshift; {\it Ref} -
Source of cluster redshift as listed below; {\it Excess} - Excess
number of sources in the central 1 Mpc (compared to the prediction)
with $1\sigma$ errors; {Centre} - log Luminosity (in erg/sec) of any
source detected within 25kpc of the cluster centre.

 [1]- \cite{NEDd}; [2]- \cite{NEDk}; [3]- \cite{NEDwg}; [4]-
\cite{NEDo}; [5]- \cite{NEDf}; [6]- \cite{NEDh}; [7]- \cite{NEDe};
[8]- \cite{NEDxj}; [9]- \cite{NEDzc}; [10]- \cite{NEDvb}; [11]-
\cite{NEDyy}; [12]- Wright, Ables \& Allen (1983)\nocite{NEDwl}; 
[13]- \cite{NEDxs}; [14]-
\cite{NEDwn}; [15]- \cite{NEDm}; [16]- \cite{NEDn}; [17]-
\cite{NEDxi}; [18]- \cite{NEDxk}; [19]- \cite{NEDs}; [20]-
\cite{NEDt}; [21]- G{\'o}mez, Hughes \& Birkinshaw
(2000)\nocite{NEDwe}; [22]- Aller, Aller \& Hughes
(1992)\nocite{NEDxn}; [23]- \cite{NEDxq}; [24]- \cite{NEDx}; [25]-
\cite{NEDzt}; [26]- Molthagen, Wendker \& Briel (1997)\nocite{NEDwr};
[27]- \cite{NEDy}; [28]- \cite{NEDza}; [29]- \cite{NEDxy}; [30]-
\cite{NEDwm}; [31]- \cite{NEDq}; [32]- \cite{NEDwv}; [33]-
\cite{NEDyn}; [34]- \cite{NEDww}; [35]- \cite{NEDyp}; [36]-
\cite{NEDyt}; [37]- \cite{NEDzg}; [38]- \cite{NEDxu}; [39]-
\cite{NEDzf}; [40]- Schade, Barrientos \& 
Lopez-Cruz (1997)\nocite{NEDzi}; [41]- \cite{NEDxy}; [42]-
\cite{NEDxc}; [43]- \cite{NEDyu}; [44]- \cite{NEDxr}; [45]-
\cite{NEDwc}; [46]- \cite{NEDwz}; [47]- \cite{NEDwx}; [48]-
\cite{NEDxh}; [49]- Abell, Corwin \& Olowin (1989)\nocite{NEDyx};
[50]- \cite{NEDxa}; [51]- \cite{NEDva}; [52]- \cite{NEDzp}; [53]-
\cite{NEDvc}; [54]- \cite{NEDwt}; [55]- \cite{NEDwu}; [56]-
\cite{NEDyq}; [57]- Cao, Wei
\& Hu (1999) \nocite{NEDu}; [58]- \cite{NEDzk}; [59]- \cite{NEDzl};
[60]- \cite{NEDwi}; [61]- \cite{NEDi}; [62]- \cite{NEDp}; [63]-
\cite{NEDr}; [64]- \cite{NEDwq}; [65]- Postman, Geller \& Huchra
(1988)\nocite{NEDxz}; [66]- \cite{NEDwo}; [67]- \cite{NEDwp}; [68]-
\cite{NEDzb}; [69]- \cite{NEDzq}; [70]- \cite{NEDyd}; [71]-
\cite{NEDxb}; [72]- \cite{NEDzs}; [73]- \cite{NEDwj}; [74]-
\cite{NEDzv}; [75]- \cite{NEDyj}; [76]- \cite{NEDwd}; [77]-
\cite{NEDzm}; [78]- \cite{NEDc}; [79]- \cite{NEDxd}; [80]-
\cite{NEDxv}; [81]- \cite{NEDxm}

\normalsize

\begin{table*}
\begin{center}
\scriptsize
\begin{tabular}{p{2.7cm}p{1.3cm}p{1.9cm}crp{1.0cm}p{0.5cm}}\label{blankfields} 
\parbox[t]{3.2cm}{Target Name}&\parbox[t]{1cm}{\vspace{2pt}Observation
RA}&\parbox[t]{1cm}{\vspace{2pt}Observation
DEC}&\parbox[t]{1cm}{\vspace{2pt}OBSID}  &\parbox[t]{1.0cm}{\raggedleft Good Exp. (ksec)}&\parbox[t]{1cm}{\vspace{2pt}Array}  &\parbox[t]{1cm}{\vspace{2pt}QSO}\\ \hline
HS0017+2116  &  00:20:10.80  &  +21:32:51.00  &  3063  &  10.0  &  ACIS-S  &  Y\\
3C9  &  00:20:25.20  &  +15:40:53.00  &  1595  &  17.5  &  ACIS-S  &  Y\\
WHDF  &  00:22:33.30  &  +00:20:55.00  &  2252  &  71.0  &  ACIS-I& \\
GSGP4X:048  &  00:57:17.10  &  -27:21:47.00  &  2242  &  10.5   &  ACIS-S&  Y$^\dag$\\
XMM1HR-3$\&$4  &  01:45:38.18  &  -04:41:24.48  &  4275,4276  &  52.1  &   ACIS-I&\\
CADIS01HFIELD  &  01:47:36.20  &  +02:20:03.30  &  2240  &  28.3  &  ACIS-I  &\\  
J0305+3525  &  03:05:47.40  &  +35:25:13.40   &  4142 &  12.3  &  ACIS-S  &  Y\\
EXTENDEDCDF-S3  &  03:31:48.79  &  -27:57:08.10  &  5019,5020  &  240.2  &  ACIS-I&\\
EXTENDEDCDF-S2  &  03:31:52.60  &  -27:41:44.92  &  5017,5018  &  219.2  &  ACIS-I&\\
EXTENDEDCDF-S4  &  03:33:01.78  &  -27:57:09.61  &  5022  &  78.7  &  ACIS-I&\\
EXTENDEDCDF-S1  &  03:33:06.10  &  -27:40:53.50  &  5015,5016  &  237.6  &  ACIS-I  &\\   
0406-244  &  04:08:51.50  &  -24:18:16.50   &  3058 &18.2    &  ACIS-S  &  Y\\
HS0818+1227  &  08:21:39.10  &  +12:17:29.00  &  3571  &  19.7  &  ACIS-S  &  Y\\
0828+193  &  08:30:53.40  &  +19:13:15.60  &  3059  &  17.4  &  ACIS-S  &Y    \\
APM08279+5255  &  08:31:41.60  &  +52:45:16.80  &  2979  &  88.3  &  ACIS-S  &  Y\\
SDSS091316+591921  &   09:13:16.60  &  +59:19:21.50  &  3034  &  9.8  &  ACIS-S  &  Y\\
QSO0910+564  &  09:14:39.30  &  +56:13:21.00  &  4821  &  22.9  &  ACIS-S  &  Y\\
BRI0952-0115  &  09:55:00.10  &  -01:30:05.00  &  5194  &  19.8  &  ACIS-S  &  Y\\
PC1000+4751  &  10:03:52.80  &  +47:36:54.30  &  4152  &  13.7  &  ACIS-S  &  Y\\
FSC10214+4724  &  10:24:34.50  &  +47:09:09.80  &  4807  &  21.4  &  ACIS-S  &  Y\\
LH-NW-4  &  10:32:06.00  &  +57:37:24.99  &  3345  &  38.3  &  ACIS-I&\\
LH-NW-6  &  10:33:22.00  &  +57:55:25.00  &  3343  &  33.5  &  ACIS-I&\\
LH-NW-5  &  10:34:02.10  &  +57:28:25.00  &  3346  &  38.1  &  ACIS-I&\\
LH-NW-9  &  10:35:16.00  &  +57:46:24.99  &  3348  &  39.4  &  ACIS-I&\\
PC\_1035+4747  &  10:38:08.20  &  +47:31:36.60  &  4154  &  8.8  &     ACIS-S  &  Y\\
SWIRELOCKMAN7  &  10:43:27.23  &  +59:10:15.07  &  5029  &  70.8  &  ACIS-I&\\
SWIRELOCKMAN1  &  10:44:46.15  &  +58:41:55.45  &  5024  &  63.7  &  ACIS-I&\\
SWIRELOCKMAN9  &  10:47:13.85  &  +59:20:06.95  &  5031  &  65.0  &  ACIS-I&\\
SWIRELOCKMAN3  &  10:48:32.77  &  +58:51:47.33  &  5026  &  68.7  &  ACIS-I&\\
Q1208+1011  &  12:10:56.90  &  +09:54:26.80  &  3570  &  10.0  &  ACIS-S  &  Y\\
HDF-N  &  12:36:49.40  &  +62:12:58.00   &  2421,3293  &  222.0  &  ACIS-I  &  \\
SDSSJ130216+003032  &  13:02:16.10  &  +00:30:32.10  &  3958  &  10.7  &  ACIS-S  &  Y\\
SDSS1306+0356JE  &  13:06:09.30  &  +03:56:43.50  &  3966  &  117.6  &  ACIS-S  &  Y\\
F864X:052  &  13:44:07.30  &  -00:28:33.00  &  2250  &  9.5  &  ACIS-S  &  Y$^\dag$\\
GROTH-WESTPHAL  &  14:17:43.60  &  +52:28:41.20  &  3305,4357,4365  &  191.3  &  ACIS-I&\\
EGS-3  &  14:20:28.00  &  +53:02:01.30  &  5845,5846  &  97.6  &  ACIS-I  &  \\
EGS-1  &  14:22:42.30  &  +53:25:37.51  &  5841,5842  &  90.6  &  ACIS-I&\\
SDSSJ144231+011055  &  14:42:31.70  &  +01:10:55.30  &  3960  &  10.8  &  ACIS-S  &  Y\\
DADDIFIELD  &  14:49:09.10  &  +09:01:36.00  &  5032,5033,5034  &  87.2   &  ACIS-I&\\
QSO1508+5714  &  15:10:02.90  &  +57:02:43.40  &  2241  &  88.5  &  ACIS-S  &  Y\\
ELAIS:N1  &  16:10:21.90  &  +54:33:36.00  &  888  &  71.9  &  ACIS-I &\\
ELAIS:N2  &  16:36:48.48  &  +41:01:45.90  &  887  &  73.1  &  ACIS-I&\\
2036-254  &  20:39:24.50  &  -25:14:30.40  &  3060  &  19.6   &  ACIS-S  &  Y\\
2048-272  &  20:51:03.40  &  -27:03:04.60  &  3061  &  17.7    &  ACIS-S  &  Y\\
\end{tabular}
\normalsize
\caption[Blank field observations]{Observations of blank fields. The fields
containing deliberately targeted QSOs are indicated - those marked $\dag$ were in the
``Extragalactic diffuse emission and surveys'' category as a search for
ROSAT identified NELGs, but no emission was seen at the target point so they
were treated as blank fields. }\label{blankfieldtable}
\end{center}
\end{table*}

\end{onecolumn}

\scriptsize  
\begin{table*}
\begin{center}
\begin{tabular}{llccrrrcc} 
Cluster&Name&RA (J2000)&DEC (J2000) & Net Counts & $F_X$&Sig&HR &$\sigma_{HR}$\\ \hline\\

3C\_295 & CXOGBA J141209.9+520419 &    14:12:09.9 &  +52:04:19.7 &       70.5 &       36.51 &       16.1 &     -0.55 &      0.22 \\
3C\_295 & CXOGBA J141127.3+521131 &    14:11:27.4 &  +52:11:31.9 &       73.4 &       21.42 &       34.0 &     -0.19 &      0.15\\
3C\_295 & CXOGBA J141135.4+521008 &    14:11:35.4 &  +52:10:08.7 &       10.7 &       4.47 &       5.2 &       1.00 &       0.00\\
3C\_295 & CXOGBA J141132.9+521103 &    14:11:33.0 &  +52:11:03.5 &       9.7 &       3.95 &       4.9 &      -1.00 &       0.00\\
3C\_295 & CXOGBA J141132.2+521116 &    14:11:32.3 &  +52:11:16.8 &       7.7 &       2.27 &       3.8 &      -1.00 &       0.00\\
3C\_295 & CXOGBA J141125.4+521047 &    14:11:25.4 &  +52:10:47.4 &       5.8 &       1.70 &       3.0 &       0.00 &       1.00\\
3C\_295 & CXOGBA J141114.4+520631 &    14:11:14.4 &  +52:06:31.1 &       60.4 &       26.16 &       21.1 &     -0.56 &      0.22\\
3C\_295 & CXOGBA J141057.4+521131 &    14:10:57.4 &  +52:11:31.1 &       27.5 &       8.16 &       13.0 &      -1.00 &       0.00\\
3C\_295 & CXOGBA J141153.0+521019 &    14:11:53.1 &  +52:10:19.9 &       25.7 &       11.29 &       10.5 &     -0.38 &      0.33\\
3C\_295 & CXOGBA J141148.3+521128 &    14:11:48.4 &  +52:11:28.8 &       10.3 &       4.37 &       4.7 &      -1.00 &       0.00\\
3C\_295 & CXOGBA J141157.8+520626 &    14:11:57.8 &  +52:06:26.2 &       30.6 &       14.29 &       9.9 &      -1.00 &       0.00\\
3C\_295 & CXOGBA J141157.3+520914 &    14:11:57.4 &  +52:09:14.0 &       9.8 &       4.30 &       3.5 &       0.00 &       1.00\\
3C\_295 & CXOGBA J141123.4+521332 &    14:11:23.4 &  +52:13:32.1 &       407.4 &       118.61 &       188.7 &     -0.60 &     0.08\\
3C\_295 & CXOGBA J141120.4+521210 &    14:11:20.5 &  +52:12:10.3 &       236.1 &       68.08 &       60.0 &     -0.25 &     0.08\\
3C\_295 & CXOGBA J141120.4+521211 &    14:11:20.4 &  +52:12:11.8 &       208.8 &       60.19 &       52.3 &     -0.75 &      0.13\\
\end{tabular}
\caption[]{Properties of sources detected in the cluster fields. The full 
sample of sources in 150 cluster fields is given in the online journal
and at http://www.sc.eso.org/$\sim$rgilmour/. Columns are (1) cluster
name given in Table \ref{Cluster_tables}, (2) full name of source
(3)\&(4) J2000 position, (5) net number of counts (6) flux in the
0.5-8keV band, $\times 10^{-15}$erg/cm$^2$/sec (assuming a spectrum
with $\Gamma=1.7$), (7) source significance as defined in Equation
\ref{SIG}, (8) counts hardness ratio, (H-S)/(H+S), where H(2-8keV) and
S(0.5-2keV) are set to 0 for detections with significance $<$ 3 in
that band (9) error on hardness ratio, where good detections exist in
both bands. }\label{sourcetable}
\end{center}
\end{table*}

\begin{twocolumn}
\newpage
\section*{Appendix: Tests for systematic errors in the pipeline}
\normalsize

\subsection*{A.1 Montecarlo simulations}\label{montecarlo}
Small fluctuations in the background level, especially in the regions
of intra-cluster emission, could cause sources to be missed by the
wavelet detection method. In order to attempt to evaluate the
detection efficiency of {\sc wavdetect} near the flux limit,
montecarlo simulations of faint sources were performed on cluster and
blank field images. The difference in the number of sources detected,
and with significance $>3$, could then be evaluated as a function of
radial position for the cluster and blank field samples.

False sources were placed in 30 cluster and 30 blank field images,
with up to 110 sources per image. The counts for each source
corresponded to a multiple (1, 1.25, 1.5 etc.) of the flux limit at
the corresponding position on the image. This procedure is not
straightforward, and may not produce accurate results as it is
difficult to simulate X-ray sources, especially those with very few
photons, because of the complex nature of the Chandra PSF. The faint
sources used in the montecarlo simulations were extracted from bright
sources at the same off-axis radius, which accounts for the off-axis
radial variation in PSF, but not any angular variation or difference
between ACIS-I and ACIS-S detectors. The small number of suitable
bright sources in the sample restricted the possible off-axis radii at
which the false sources could lie.

There is a small deviation between the detection rates in cluster and
blank fields in the central 100 arcsec only. Surprisingly rather than
the detection rate decreasing for the cluster fields, it increases
slightly for the blank fields. This is most likely due to the problems
with producing accurate input sources, as described above. Sources at
the flux limit were around 15\% more likely to be detected in the
blank fields in the central 100 arcsec, but this difference decreases
rapidly as the source flux increases. Combining the results for the
cluster and blank fields with the Log N($>S$) - Log S distribution
gives an estimate of the number of sources missed in the cluster
fields relative to the blank field prediction. This is found to be
$\sim 0.12$ sources per cluster field in the central $25\arcsec
-100\arcsec$. Very few sources are expected to be missed in the
central $25\arcsec$ due to the low area and high flux limit. The
maximum errors due to missed sources result in a $\lesssim 1 \sigma$
change in the results.

\subsection*{A.2 Cosmic variance in small samples}\label{smallsamples}

The error calculations used here are based on the Poissonian errors on
the sources detected, assuming that they are randomly
distributed. However large scale structure may give rise to
significantly larger errors in small samples. The clustering of X-ray
sources appears to be stronger in low flux sources than in high flux,
and can give rise to significant field to field variations (see for
example \citeauthor{Yang2} \citeyear{Yang2}, \citeauthor{Mullis}
\citeyear{Mullis} and \citeauthor{Basilakos} \citeyear{Basilakos}). To
check the magnitude of this effect in this survey, subsamples of 1 to
5 blank fields were chosen at random. Figure \ref{blank_variance2}
shows the ratio between the observed variation in the Log N($>$S) -
Log S and the expected variation from Poissonian errors, for 1000
subsamples of each size. At $10^{-14}$erg cm$^{-2}$ s$^{-1}$ the
number of sources is small and the Poissonian errors completely
explain the variance between the sub-samples. At $10^{-14.5}$erg
cm$^{-2}$ s$^{-1}$ the variation in individual fields is $20\%$ larger
than that expected from the Poissonian errors, which is attributable
to AGN clustering. This effect decreases as the sample size is
increased, and for samples of 5 fields the variation is only slightly
above the expected value; samples of this size are therefore
sufficient to largely average out the effects of large scale
structure. At $10^{-15}$erg cm$^{-2}$ s$^{-1}$ there are few blank
fields available so only the variance for individual fields is
shown. This is $\sim 10\%$ higher than expected from Poissonian
analysis, indicating that there is some effect due to large scale
structure in faint sources too. In taking small samples of fields
it is therefore advisable to use at least 5 fields in order to ensure
that the stated errors are not underestimated due to clustering.

\begin{figure}
\begin{center}
\includegraphics[trim=0.5cm 0cm 0cm 0cm,clip,width=\columnwidth]{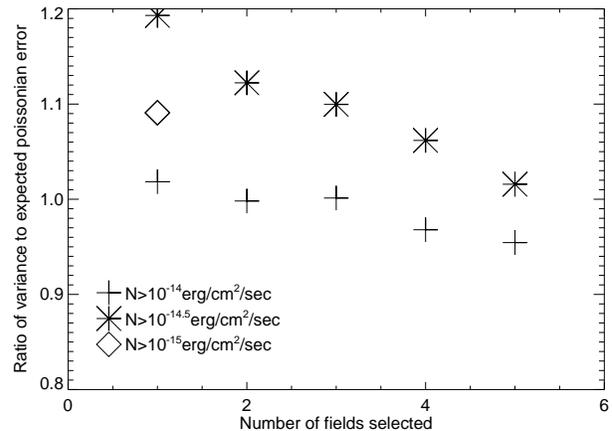}
\caption[The variance in subsamples of blank fields, compared to that
predicted from a random distribution.]{The average ratio of actual to
expected (Poissonian) variance in N($>S$)for sources detected in the
0.5-8 keV band in 1000 sub samples of blank fields. Sub-samples of 1 -
5 fields are investigated at three flux values. This result has two
limiting factors: Firstly the subsamples were picked from the same
parent population (in particular at $N > 10^{-15}$erg cm$^{-2}$
s$^{-1}$ only 17 blank fields were available) so the overlap between
subsamples reduces the observed variance for moderate sample
sizes. Secondly the values can be $<1$ as the Poissonian errors were
calculated from the mean field value, whereas in reality some fields
are larger so have more sources and smaller
errors.}\label{blank_variance2}
\end{center}
\end{figure}

\subsection*{A.3 Blank field results}\label{Check_blanks}
The results for the 44 blank fields were checked to ensure that the
method and pipeline worked correctly. The blank field Log N($>$S) -
Log S was compared to the literature, and the radial distribution of
blank field sources was compared to the pipeline prediction for the same
fields.

\begin{figure}
\begin{center}
\includegraphics[clip,width=\columnwidth]{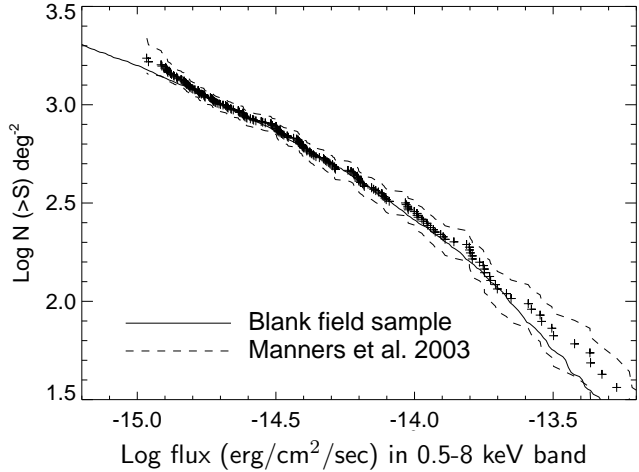}
\caption[]{The blank field Log N($>$S) - Log S plot compared to that from the
 ELAIS fields of \cite{Manners}. The solid line is the data from this paper. 
The crosses mark the data points and the dashed lines the 1$\sigma$ errors 
from the ELAIS fields.}\label{JCM_blank}

\end{center}
\end{figure}

The blank field Log N($>S$) - Log S distribution was compared with
that derived by \cite{Manners} from the ELAIS fields, as shown in
Figure \ref{JCM_blank}, and agrees to well within the 1$\sigma$ error
bars. As with most blank field surveys, the ELAIS sample will be
affected by sample variance as it only covers 2 Chandra
fields. Unfortunately, all other blank field Chandra surveys calculate
the Log N($>S$) - Log S for the 0.5-2 keV and 2-8 keV bands
independently, so cannot be compared to the pipeline results directly.


\begin{figure}
\begin{center}
\includegraphics[width=\columnwidth]{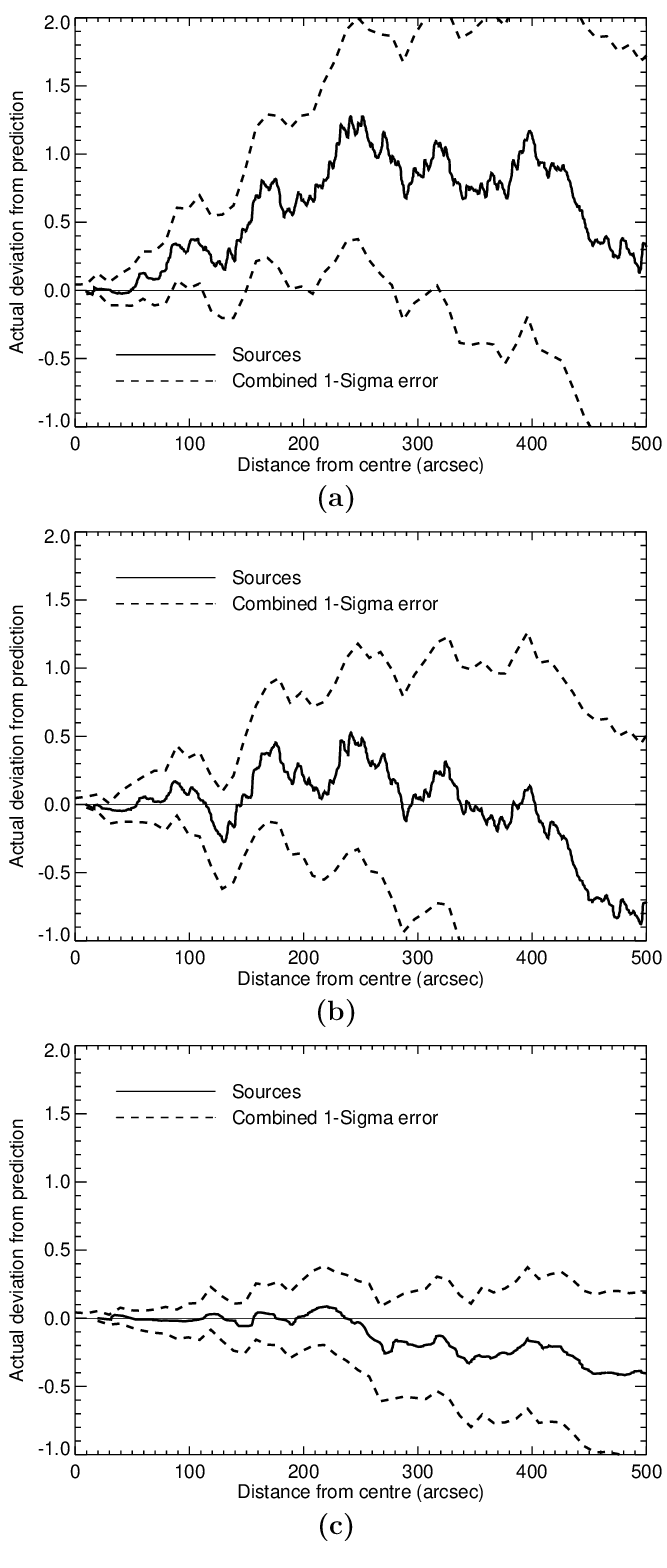}
\caption[Radial distribution and prediction for blank fields]{The radial 
prediction and actual distribution of sources in blank fields, with
$1\sigma$ errors, for (a) all sources in all 44 blank fields; (b) all
sources in 40 blank fields, excluding the four fields with very deep
exposures, which only cover two regions of the sky. These two regions
bias the sample as they contribute a significant fraction of the
sources and are both overdense due to cosmic variance; (c) sources
brighter than $10^{-14}$erg cm$^{-2}$sec$^{-1}$, in all 44 blank
fields.  }\label{blank_all_radial}
\end{center}
\end{figure}

The combined radial distribution for all blank field sources is shown
in Figure \ref{blank_all_radial}(a). In the galaxy cluster
observations the cluster is not generally placed at the centre of the
detector array. Therefore, in order to reproduce the method used for
the cluster fields, the radial distribution for the blank fields was
measured from a point on the detector which corresponded to the
cluster centre, in a randomly selected cluster observation. There is a
slight excess of sources in the central 200\arcsec compared to the
prediction. Although this is not very significant it does correspond
to around one extra source per average field. However, the blank field
sample is dominated by the four deepest observations, which actually
only cover two regions of sky - the Chandra/Hubble Deep Field North
and the Extended Chandra Deep Field South. These two regions are a
factor of two deeper in exposure time than the average blank fields,
and account for 13 per cent of the total sources, and 35 per cent of
the fainter sources ($<10^{-14.5}$erg cm$^{-2}$sec$^{-1}$). Because these
two regions contain so many sources, a small variation in the number
of sources carries more weight than for the shallower blank fields.
The effect of cosmic variance is therefore amplified in Figure
\ref{blank_all_radial}(a), as explained in Section A.2.

Figure \ref{blank_all_radial}(b) shows the same radial distribution,
but excluding the two regions of sky covered by the four deepest
observations, to limit the effects of cosmic variance. The remaining
blank fields have far less range in exposure times, and as expected
the excess in the radial distribution seen in Figure
\ref{blank_all_radial}(a) disappears. This sample of blank fields has a 
similar range of exposure times to the cluster sample. Figure
\ref{blank_all_radial}(b) therefore shows that the prediction is
accurate so long as the source counts are not dominated by a few
fields. Section A.2 gives a more quantitative analysis of this
problem.

Figure \ref{blank_all_radial}(c) shows the radial distribution for all
sources brighter than $10^{-14}$erg cm$^{-2}$sec$^{-1}$. Again the
source distribution matches the prediction well. At this flux level
all blank fields contribute to the source counts, and there is no
problem due to cosmic variance. Figures \ref{blank_all_radial}(b) and
(c) can be compared to Figure \ref{radl3}(a)-(c) to show that the
excess seen in the cluster fields is real.

As a final check the total excess or deficit of sources (over the full
radius) compared to the prediction, was examined for each blank
field. There was no noticable correlation between the deviation from
the prediction and factors such as exposure time, or whether a field
was merged or not. There was a small and insignificant correlation
with ACIS array, as explained in Section A.4 below.

\subsection*{A.4 The CCD array}\label{ccdarray}

To check for systematic offsets between the pipeline results for
fields observed with the ACIS-I and ACIS-S detectors, the radial
distribution and Log N($>S$) - Log S distributions for blank fields
observed with each detector were compared. It is also desirable to
check for differences between the true blank fields and those which
targeted high redshift QSOs. Unfortunately these cannot be done
independently as the 22 high redshift QSO fields were all observed with
ACIS-S, and the 22 true blank fields were observed with ACIS-I.

Figure \ref{blank_is_LognLogs} shows the difference between the Log
N($>$S) - Log S distributions for the `true' blank and QSO fields.
The QSO fields (ACIS-S) have a Log N($>$S) - Log S distribution that
is around 1$\sigma$ {\it lower} than the `true' blanks (observed with
ACIS-I), so there are no significant extra sources in the high
redshift QSO fields, and they are valid blank fields.

\begin{figure}
\begin{center}
\includegraphics[trim=0.5cm 0cm 0cm 0cm,clip,width=\columnwidth]{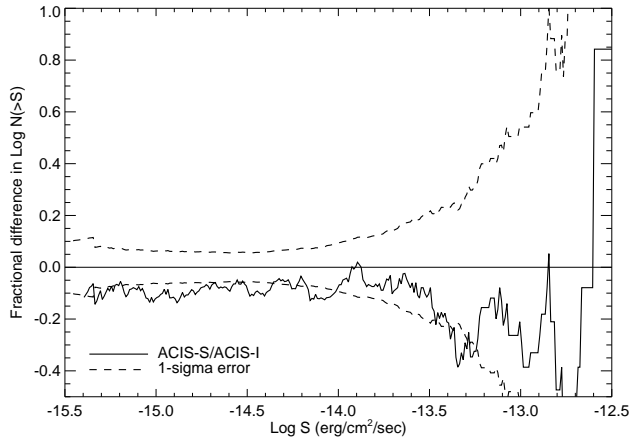}
\caption[Log N($>$S) - Log S plots for blank fields observed with
ACIS-I and ACIS-S.]{Fractional difference between the Log N($>$S) -
Log S plots for ACIS-S and ACIS-I blank fields. The 1$\sigma$ errors
are shown relative to the ACIS-I line, and are found by combining the
errors on the two Log N($>$S) - Log S distributions. The ACIS-S
distribution is around 1$\sigma$ lower than the ACIS-I distribution at
all fluxes.}\label{blank_is_LognLogs}
\end{center}
\end{figure}

The 1$\sigma$ offset between the ACIS-I and ACIS-S (which correspond
to the QSO and `true' blank fields) is of low significance, but it is
worth checking that it is not a systematic error. The small difference
in the source size between ACIS-I and ACIS-S images cannot account for
the 1$\sigma$ variation (Section \ref{Model}). As the offset is also
seen at higher fluxes, where the sky area is given by the total area
of the detector, errors in the calculation of $S_{min}$ also cannot
explain the difference. In addition the flux calibration between
ACIS-I and ACIS-S is accurate to within $\sim 5\%$\footnote{see
details in
\\ http://cxc.harvard.edu/cal/docs/cal\_present\_status.html} , whereas
a $10\%$ offset would be required to change the Log N($>$S) - Log S by
1$\sigma$. Finally, the effect of sources overlapping at the edges of
the images (ACIS-I observations have more large sources) is minimal,
even for the deepest fields.

Instead, the small offset between ACIS-I and ACIS-S number counts is
most likely to be just due to the high statistical variance between
the fields.  The 44 blank fields were split into two equal subsamples
and the difference in $N(S > 10^{-14}$erg cm$^{-2}$ s$^{-1})$ was
computed. This was repeated for 1000 randomly chosen
subsamples. Figure
\ref{blank_variance} shows that the difference between the ACIS-I and
ACIS-S samples is fully consistent with randomly chosen samples of
blank fields. The chance of getting a difference of $>23$ sources
between the two samples is $\sim$37 per cent, which is in full
agreement with the size of the $1\sigma$ error bars in Figure
\ref{blank_is_LognLogs}.

\begin{figure}
\begin{center}
\includegraphics[trim=0.5cm 0cm 0cm 0cm,clip,width=\columnwidth]{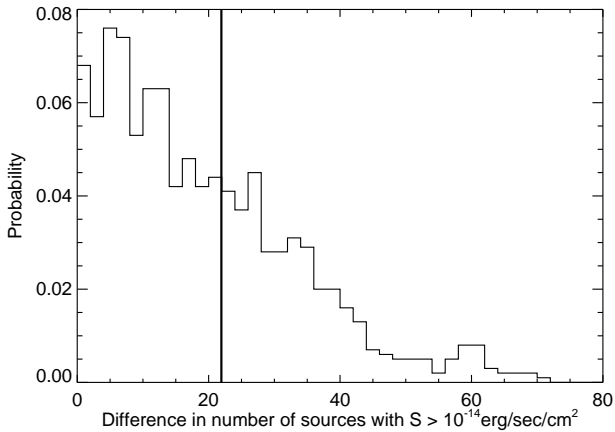}
\caption[ ]{ The difference between the total number of sources $> 10^{-14}$erg
cm$^{-2}$sec$^{-1}$ when the 44 blank fields are split into two equal
subsamples, for 1000 randomly chosen subsamples.  The thick vertical
line shows the value when the sample is split into the 22 ACIS-I and
22 ACIS-S fields (see Figure \ref{blank_is_LognLogs}). This value is
consistent with 22 randomly chosen fields (the probability of having
two such different values randomly is 37 per cent) and so there is no evidence
of a systematic offset in the number of sources detected in ACIS-I and
ACIS-S blank fields.  }
\label{blank_variance}
\end{center}
\end{figure}

The radial distributions for the ACIS-I and ACIS-S fields were also
compared. Comparing the results for ACIS-I and ACIS-S blank fields
against the prediction from the blank field Log N($>$S) - Log S, the
ACIS-S fields end 1$\sigma$ below the prediction in agreement with
Figure \ref{blank_is_LognLogs}.  Otherwise both distributions are flat
to within the errors, and there is no link between over or under
prediction and radius.

\end{twocolumn}

\label{lastpage}

\end{document}